\documentclass[10pt,journal,cspaper,compsoc]{IEEEtran}
\ifCLASSOPTIONcompsoc
\else
\fi
\ifCLASSINFOpdf
\else
\fi

\usepackage{amsthm}
\newtheorem{theorem}{Theorem}
\usepackage{graphicx, epsfig, amssymb, amsmath}
\usepackage{subfig}
\usepackage[linesnumbered,ruled,vlined]{algorithm2e}
\usepackage{hyperref}
\usepackage{caption}
\usepackage{ragged2e}
\usepackage{lipsum}
\usepackage[font={footnotesize,it}]{caption}

\begin{document}
%
\title{Security Pricing as Enabler of Cyber-Insurance\\ \emph{A First Look at Differentiated Pricing Markets}}
%
%
%
%

\author{Ranjan~Pal,~\IEEEmembership{Member,~IEEE,}
           \{Leana~Golubchik, Konstantinos Psounis\},~\IEEEmembership{Senior Member,~IEEE,}
        ~Pan~Hui,~\IEEEmembership{Senior Member,~IEEE}
\thanks{R.Pal is with the department
of Computer Science, University of Southern California,
CA, USA. E-mail: \emph{rpal@usc.edu.}}
\thanks{L. Golubchik is jointly with the departments of Computer Science and Electrical Engineering, University of Southern California. E-mail: \emph{leana@usc.edu}.}
\thanks{K. Psounis is jointly with the departments of Electrical Engineering and Computer Science, University of Southern California. E-mail: \emph{kpsounis@usc.edu}.}
\thanks{P. Hui is jointly with the department of Computer Science and Engineering, Hong Kong University of Science and Technology, and Telekom Innovation Labs (Y-Labs), Berlin. E-mail: \emph{pan.hui@telekom.de}.}}

\markboth{Submitted to IEEE Transactions on Dependable and Secure Computing}%
{Shell \MakeLowercase{\textit{et al.}}: Bare Demo of IEEEtran.cls for Computer Society Journals}
%


\IEEEcompsoctitleabstractindextext{%
\begin{abstract}
\justifying
Despite the promising potential of network risk management services (e.g., cyber-insurance) to improve information security, their deployment is relatively scarce, primarily due to such service companies being unable to guarantee profitability. As a novel approach to making cyber-insurance services more viable, we explore a symbiotic relationship between security vendors (e.g., Symantec) capable of price differentiating their clients, and cyber-insurance agencies having possession of information related to the security investments of their clients. The goal of this relationship is to (i) allow security vendors to price differentiate their clients based on security investment information from insurance agencies, (ii) allow the vendors to make more profit than in homogeneous pricing settings, and (iii) subsequently transfer some of the extra profit to cyber-insurance agencies to make insurance services more viable. 

\noindent In this paper, we perform a theoretical study of a market for differentiated security product pricing, primarily with a view to ensuring that security vendors (SVs) make more profit in the differentiated pricing case as compared to the case of non-differentiated pricing. In order to practically realize such pricing markets, we propose novel and  \emph{computationally efficient} consumer differentiated pricing mechanisms for SVs based on (i) the market structure, (ii) the communication network structure of SV consumers captured via a consumer's \emph{Bonacich centrality} in the network, and (iii) security investment amounts made by SV consumers. We validate our analytical model via extensive simulations conducted on practical SV client network topologies; main results show (through those simulations) that (a) a \emph{monopoly} SV could improve its profit margin by upto $\approx$ 25\% (based on the simulation setting) by accounting for clients' investment information and network locations, whereas in an \emph{oligopoly} setting, SVs could improve their profit margins by upto $\approx$ 18\%, and (b) differentiated security pricing mechanisms are fair among SV consumers with respect to the total investment made by a consumer. To the best of knowledge, the proposed differentiated pricing framework is the first of its kind in the security products domain, and is generally applicable to usecases beyond the one investigated in this work.  

\end{abstract}

}

\maketitle

\IEEEdisplaynotcompsoctitleabstractindextext

%
\IEEEpeerreviewmaketitle

\section{Introduction}\label{sec-intro}
The infrastructure, the users, and the services offered on computer networks today are all subject to a wide variety of risks. These risks include distributed denial of service attacks, intrusions of various kinds, eavesdropping, hacking, phishing, worms, viruses, spams, etc. Network users (both individuals and organizations) have traditionally resorted to antivirus and anti-spam software, firewalls, intrusion-detection systems (IDSs), and other add-ons to reduce the likelihood of being affected by threats. Currently, a large industry (companies like \emph{Symantec, McAfee,}) as well as considerable research efforts are centered around developing and deploying methods to detect threats and anomalies in order to protect the cyber infrastructure and its users from the negative impact of the anomalies.

\subsection{Technology Drawbacks and Risk Management}
Inspite of improvements in risk protection techniques over the last decade due to hardware, software and cryptographic methodologies, \emph{it is impossible to achieve perfect/near-perfect cyber-security protection} \cite{amr}\cite{leb}. The impossibility arises due to a number of reasons (see \cite{amr} for details): (i) scarce existence of sound technical solutions, (ii) difficulty in designing solutions catered to varied intentions behind network attacks, (iii) misaligned incentives between network users, security product vendors, and regulatory authorities regarding each taking upon them a proper liability to protect the network, (iv) network users taking advantage of the positive security effects 
generated by other user investments in security, in turn themselves not investing in security and resulting in the free-riding problem, (v) customer lock-in and first mover effects of vulnerable security products, (vi) absence of popular vulnerability markets, e.g., settings where vulnerabilities are traded for money, (vii) difficulty measuring risks resulting in challenges to designing pertinent risk removal solutions, (viii) the problem of a lemons market \cite{lmr}, whereby security vendors have no incentive to release robust products in the market, (ix) liability shell games played by product vendors, and (x) user naiveness in optimally exploiting feature benefits of technical solutions. In view of these inevitable barriers to 100\% risk mitigation, the need arises for alternative methods of risk management in cyberspace. In this regard, some security researchers in the recent past have identified \emph{cyber-insurance} as a potential tool for effective risk management. 

\emph{Cyber-insurance} is a risk management technique via which network users transfer their risks to an insurance company (e.g., ISP, cloud provider), in return for a fee, i.e., the \emph{insurance premium}. Researchers vouching for the use of cyber-insurance believe that cyber-insurance would lead to the design of insurance contracts that would shift appropriate amounts of self-defense liability on the clients, thereby making the cyberspace more robust. Here the term `self-defense' implies the efforts by a network user to secure their system through technical solutions such as anti-virus and anti-spam software, firewalls, using secure operating systems, etc. Cyber-insurance can also potentially be a market solution that can align with economic incentives of cyber-insurers, users (individuals/organizations), policy makers, and security software vendors, i.e., the cyber-insurers will earn profit from appropriately pricing premiums, security risks will be appropriately shared among network users and the latter will seek to hedge potential losses by jointly buying insurance and investing in self-defense mechanisms, the policy makers would ensure the increase in overall network security, and the security software vendors could go ahead with their first-mover and lock-in strategies, and at the same time experience an increase in their product sales via forming alliances with cyber-insurers. 

\subsection{Research Motivation}
We motivate our work on differentiated security pricing by first raising the issue of moderate/unviable cyber-insurance markets, and then introducing the concept of price differentiating security products, and how it might, as an idea, resolve the problem of unviable insurance markets. \emph{We emphasize here that the use case of cyber-insurance markets is just one instance where the concept of price differentiating security products might be applicable.}
\subsubsection{Moderate Cyber-Insurance Markets}
The total cyber-insurance business currently amounts to US\$ 2 billion, whereas the total cost of security breaches to the global economy amounts to a whopping US\$ 445 billion \cite{mthr}. There are presently over 30 insurance carriers in the United States offering cyber-insurance contracts, and as of 2015, these carriers report significant yearly increases (up to approximately 25\%) in their client base, which has primarily been corpororate organizations (ref. \emph{Betterley Report}). However, the current inability to bridge the US\$ 443 billion gap could be overcome if the cyber-insurance business were to spread to individuals rather than only industries and organizations. A major roadblock is the potential inability of the cyber-insurers to make strictly positive profit at all times \cite{rpinfr}\cite{leb}\cite{ssfw}, primarily due to the high degree of non-transparency of risk information between the insurer and the insured in a networked setting. \emph{Hence, the need of the hour is to design alternative mechanisms that allow cyber-insurers to make positive profit at all times \textbf{without relying on the degree of information transparency between the insurer and the insured}, and in-turn enable wide-spread adoption of cyber-insurance products amongst individuals human users, industries, and organizations.} 

\noindent \textbf{Our Main Idea -} We envision price differentiation of security products amongst SV consumers to be a mechanism that will make cyber-insurance markets more viable, without it requiring information on insurance parameters such as premium or coverage. 
\subsubsection{A Need for Price Differentiating Security Products}
Security products (e.g., antivirus software) are one of the main sources of protection for Internet users to prevent their communication devices from being hacked. An important factor on the basis of which cyber-insurance companies charge premiums to their clients (users) is the latters' expected risk value, which in turn is a function of, along with other parameters, the robustness of security protection adopted by the clients. In practice, most users do not take advantage of the full power of a security product, either due to the ignorance of using the product effectively, or being willfully careless. In both cases, these users, along with the careful and security knowledgeable users, usually pay the same price for a security product available in the market. Our idea in this paper is to make SVs charge/price users (consumers) \emph{heterogenously} but \emph{proportionately}, based on their security behavior. \emph{We hypothesize, based on price differentiation principles in economics, that in doing so the security vendors would make more profit compared to their usual uniform pricing mechanism.} As a result, they could symbiotically transfer a portion of their profit amounts to cyber-insurers in return for the latter providing the vendors with (i) information about consumer security behavior that they collect, and (ii) vendor lock-in privileges. The making of extra profit by the cyber-insurers in this symbiotic environment can improve their chances of ensuring market viability. 

\noindent \textbf{Goal:} \emph{In this paper, our two-fold goal is to (i) prove our hypothesis that proportionately fair heterogenous security product pricing among Internet users generates more profits for SVs compared to uniform pricing, and (ii) realize a practical, computationally efficient way to achieve price heterogeneity.}

\subsection{Research Contributions} \label{sec-rc}
We make the following contributions in this paper. 
\begin{enumerate}
\item We propose a pricing environment consisting of security vendors (SVs) and their clients, and mathematically model the vendor-client interaction mechanism that accounts for client security investment, and the positive externalities caused due to them (see Section 2). 
\item We propose a static and heterogenous product pricing mechanism for SV clients based on the client (consumer) logical network and their security investment amounts. Our proposed static pricing mechanism also subsumes the uniform and binary pricing (different prices for two different user types) mechanisms as special cases of the general heterogenous pricing mechanism. The latter two pricing mechanisms are of use when the SV might be constrained in practice (e.g., due to policy issues) to adopt differentiated pricing schemes on security products. Our proposed pricing mechanisms are based on \emph{Stackelberg games} \cite{ft}. We show in theory that there always exists a unique Nash equilibrium value of the SV prices and investments of their corresponding consumers, for the pricing game entailed by the mechanisms. In addition, we also show that despite the existence of a unique Nash equilibrium, the design of an optimal binary pricing mechanism is an NP-Hard problem, for which we design an efficient randomized-approximation algorithm. Finally, using spectral graph theory, we also derive tight bounds on the ratio of the profit margins for a monopoly SV, with and without taking into account network externalities (see Section 3). 
\item We conduct an extensive numerical evaluation study for monopoly and oligopoly SV settings to highlight the effects of consumer overlay network on SV heterogenous pricing outcomes. Specifically, for practical real world topologies like \emph{scale free graphs} and \emph{trees} (more details in Section 4), we show that (i) the per-unit product price charged by an SV is proportional to the Bonacich centrality of consumers in their overlay network, thus obeying the results obtained in theory from Section 3, and (ii) the \emph{total cost} incurred by every consumer (network user) in security investments is nearly equal, and amounts to a constant that is independent of the underlying network topology. The latter point implies consumer fairness (a notion similar to network neutrality) because no matter how a consumer is placed in an overlay network, he pays the same total amount in security investments as any other consumer in the network, even though his per-unit security investment price charged by an SV is proportional to the amount of positive externalities he generates via his investments. Via simulation results, we also observe that (i) a \emph{monopoly} SV could improve their current profit margins by $\approx$ upto 25\% (based on the simulation setting) by accounting for client location information in the consumer network and his investment information, whereas (ii) in an \emph{oligopoly} setting, SVs could increase their current profit margins by $\approx$ upto 18\% (see Section 4).
\end{enumerate}

As a final comment, to the best of our knowledge, the proposed differentiated pricing framework is the first of its kind in the security products domain, and is generally applicable to usecases beyond the one investigated in this work.  

\subsection{Related Efforts}
As mentioned above, with respect to the pricing economics of security products, ours is the first work on differentiated security product pricing for consumers interacting via a network. In this section, we first provide a brief description of existing research related to non-homogenous pricing in a network, where externalities and the network structure are accounted for. We then state the differences of this work with our preliminary results \cite{rpinfr1}. 

\noindent \textbf{Network and Investment Dependent Pricing -} Given a set of prices, our model takes the form of a network game among interacting consumers. Recent papers that study such games include \cite{ballester2006s}\cite{bramoulle2007public}\cite{corbo2007importance}\cite{galeotti2009influencing}.  A key modeling assumption in \cite{ballester2006s}\cite{bramoulle2007public}\cite{corbo2007importance}, that we also adopt in our setting, is that the payoff function of a consumer takes the form of a linear-quadratic function. Ballester et al. \cite{ballester2006s} were the first to note the linkage between Bonacich centrality and Nash equilibrium outcomes in a single stage game with local payoff complementarities. Our characterization of optimal prices when the monopolist can perfectly price differentiate is reminiscent of their results, since prices are inherently related to the Bonacich centrality of each consumer. \emph{However, both the motivation and the analysis are quite different}, since ours is a two-stage game, where a monopolist chooses prices to maximize revenue subject to equilibrium constraints. Also, \cite{bramoulle2007public} and \cite{corbo2007importance} study a similar game and interpret their results in terms of public good provision. A number of recent papers \cite{galeotti2010network} \cite{sundararajan2007local} make the assumption of limited knowledge of the logical network structure, i.e., they assume that only the degree distribution is known, and thus derive optimal pricing strategies that depend on this first degree measure of externalities of a consumer. In our model, we make the assumption that the monopolist has complete knowledge of the logical network structure and, thus, obtain qualitatively different results: \emph{the degree is not the appropriate measure of influence due to externality, but rather prices are proportional to the Bonacich centrality of the agents.} On the technical side, note that assuming more global knowledge of the network structure increases the complexity of the problem in the following way: if only the degree of a consumer is known, then essentially there are as many different types of consumers as there are different degrees. This is no longer true when more is known: then, two consumers of the same degree may be of different type because of the difference in the characteristics of their neighbors, and therefore, optimal prices charged to consumers may be different. \emph{Also, as major differences from the above mentioned works, (i) for the monopoly setting, we derive tight bounds on the ratio of the profit margins for an SV, with and without taking into account network externalities, and (ii) We study (via simulations) network price differentiation in oligpolistic markets consisting of multiple competing SVs.} 

\noindent \textbf{Differences with Preliminary Results  \cite{rpinfr1} -} In a previous effort \cite{rpinfr1} related to this paper, we had formalized the differentiated pricing problem for network users and proved it to be NP-Hard (for the binary pricing case). In this paper, we surmount the computational challenges of the binary pricing problem and propose an approximation algorithm for that problem (see Section 3.3, Theorem 5 and its proof.). Also, in contrast to \cite{rpinfr1}, in this paper we (i) validate the effectiveness of our proposed algorithm via extensive simulations run on monopoly and competitive security vendor markets, for practical network topologies (see Section 4), and (ii) mathematically characterize the difference (via ratio bounds) in profit for SVs with and without price differentiation (see Section 3.2, Theorem 3 and its proof), for general topologies.

\section{System Model} \label{sec-setup}
In this section we propose our seller-buyer system model. We first, we qualitatively describe the general environment comprising SVs (the sellers), their clients (the buyers), and the agencies (e.g., cyberinsurance firms) that might benefit from SV pricing mechanisms. We then describe the SV pricing rationale. Finally, we analytically define the seller-buyer interaction framework. A summary of main notation is shown in Table 1.  

\begin{table}
\footnotesize
\centering
\begin{tabular}{|c|c|}
\hline
Symbol & Meaning\\
\hline
$u_{i}(\cdot)$ & utility of user (consumer) $i$ in consumer-seller model\\
$N$ & number of consumers of an SV \\
$h_{ij}$ & externality effect of user $j$ on user $i$\\
$x_{i}$ & amount of self-defense goods consumed by user $i$\\
$G$ & matrix representing externality values between user pairs\\
$\overrightarrow{x_{-i}}$ & vector of self-defense amounts of users apart from $i$\\
$B(\cdot)$ & Bonacich centrality vector of users in a logical network\\
$c$ & constant marginal manufacturing cost to SV\\
$p_{i}$ & price per unit of self-defense good consumed by $i$\\
\hline
\end{tabular}
\caption{Notation}
\label{tab:template}
\end{table}

\subsection{General Setting}
We consider a system where potentially multiple security vendors exists in a market. We assume that all network users have access to some paid security software, e.g., antivirus software, developed by a given security vendor \footnote{It is a well known fact that many Internet users use free pirated versions of anti-virus software, or sample free versions. For simplicity, we do not consider this issue in the paper. However, we believe that this issue can be resolved by regulatory agencies via appropriate policies and mandates so that every user is officially associated with a particular security vendor.}. We assume that SV clients form a logical network (e.g., via Gmail, Facebook, Twitter, etc.). These networks are the most common substrate for facilitating social engineering attacks that are currently amongst the most common cyberattack methods in practice \cite{pc1}. However, our work is also applicable to non-logical network structures. 

With respect to a cyber-insurance setting (the example setting for our paper motivating price differentiation by SVs), the SVs can form a symbiotic relationship with ISPs, which can act as a cyber-insurance agency. In view of recent recommendations made by an FCC advisory committee\footnote{FCC chairman Julius Genachowski spoke at the Cybersecurity Bipartisan Policy Center on the role of ISPs in improving cybersecurity  (2012).}, ISPs have committed to taking steps to combat cyber-security threats (e.g., enforcing antibot conduct code, executing the ability to monitor, trace, analyse, and block traffic without violating privacy rules, and educating Internet users to access the web safely. Thus, we envision a future where SVs can work in collaboration with ISPs to make cyber-space secure. The ISPs can act as cyber-insurers, where a client locks-in with particular SV through an ISP providing service to the client. This could happen if while signing an Internet agreement with the ISP, the client is required to buy security products by a particular vendor in a symbiotic business relationship with an ISP. 

\subsection{SV Pricing Rationale} \label{sec-price}
We consider SVs adopting a heterogenous product pricing mechanism that is based on the logical communication network of its consumers and their corresponding security investments. The motivation for heterogenous pricing manner is for an SV to make more profit from its security products compared to that obtained from uniform pricing, through extra client information obtained from ISPs. In addition, it would also seem fair to price network users heterogenously due to the differing amounts of positive externalities they generate through their security investments. With a uniform pricing mechanism, some users pay more compared to the externality they generate, and vice versa, thereby leading to the existence of free-riders in the network. One way for SVs to envision a free-rider free network is to deploy a heterogenous pricing scheme for their clients. A thing to note here is that price heterogeneity or discrimination in general might not go down well with consumers, primarily raising network neutrality concerns. However, SVs can get topology and security investment related information about their clients from ISPs (and hence proportionately estimate client generated externalities) through disclosure agreements signed between the ISP, the SV, and their clients, as part of a mandate imposed by the government \cite{rabohme}. In doing so, the SVs can convince consumers of a heterogenous but fair pricing mechanism in operation that will raise fewer eyebrows. 

\subsection{Seller-Buyer Interaction} \label{sec-model}
To mathematically illustrate the buyer-seller interaction, for simplicity of analysis, in this section we only consider a monopoly setting with a single SV serving clients forming a logical network.  In this regard, we assume that a monopolistic SV (seller) has $N$ clients (consumers), connected via a logical network and using self-defense products manufactured by the SV. 
Each consumer $i\,\epsilon\,N$ has a utility function, $u_{i}(\cdot)$, given as 
\begin{equation}
u_{i}(x_{i}, \overrightarrow{x_{-i}}, p_{i}) = \alpha_{i}x_{i} - \beta_{i}x_{i}^{2} + x_{i}\cdot\sum_{j}h_{ij}x_{j} - p_{i}x_{i},
\end{equation}
where $x_{i}$, a continuous variable, is the amount of self-defense/self-protection features consumed or invested in by user $i$, $\overrightarrow{x_{-i}}$ is the vector of self-defense investments of users other than $i$, and $p_{i}$ is the price charged by the SV to user $i$ per unit of self-defense investment consumed by $i$.  In reality, SV products are bundled in a package which is priced as a single item. However, every bundled package has a number of features which different users use differently, and the effectiveness of a user's security protection would depend on how he uses those features \cite{pc1}. In that light, we assume that $x_{i}$ is continuous. Here $p_{i}$ is the equilibrium market price set by the SV (in case of a competitive setting, the equilibrium price is set after competing with other SVs in the security product business.).  $\alpha_{i}, \beta_{i}, h_{ij}$ are constants. $\alpha_{i}, \beta_{i}$ are constants associated with a user's individual investments $x_{i}$, and 
$h_{ij}$ is the amount of externality user $j$ exerts on user $i$ through his per unit investments. Here $h_{ij} \ge 0$ and $h_{ii} = 0,\forall i$. $x_{i}$  is assumed to be continuous for analysis tractability reasons. The first and second term in the utility function denote the utility to a user solely dependent on his own investments, the third term is the positive externality effects of investments made by other users in the network on user $i$, and the fourth term is the price user $i$ pays for consuming $x_{i}$ units of self-defense goods manufactured by the SV. We assume here that $x_{i}$ is bounded. \emph{It is well known that the quadratic nature of the utility function allows for a tractable analysis and offers a nice second-order approximation to higher order concave payoffs, while preserving the properties of utility functions for risk-averse users. The insight into why quadratic functions serve as a good approximation to higher order concave functions is related to the point in the function curve when the law of diminishing returns starts to hold. For different cost functions the point is different, but inherently it exists on every curve. So the trends one observes by analyzing (or simulating) general concave costs is the same as those observed by analyzing (simulating) quadratic concave functions.}

The SV accounts for the strategic investment behavior (after it would have set its prices) of users it provides service to, and decides on an optimal pricing scheme that arises from the solution to the following unconstrained optimization problem. 
\[max_{\overrightarrow{p}}\sum_{i} p_{i}x_{i} - cx_{i},\]
 where $\overrightarrow{p}$ are the vectors of prices charged by the SV to its consumers, and $x_{i}$ is the amount of self-defense goods consumed by consumer $i$ after the SV sets its prices. $c$ is the constant marginal cost to the SV to manufacture a unit of  any of its products. For the analysis that follows in the paper, we will assume for all $i$, (i) $2\beta_{i} > \sum_{j\,\epsilon\,N}h_{ij}$ and (ii) $\alpha_{i} > c$. Assumption (i) implies that the concavity of user utility functions and that the optimal investment level of network users are bounded, and assumption (ii) implies that all network users purchase a positive amount of security product manufactured by the SV. 


\section{The SV Pricing Mechanism} \label{sec-price}
Our goal in this section is to (i) study and analyze optimal pricing mechanisms for SVs and the corresponding security investments of their clients (consumers), and (ii) to draw potential relationships between the structure of the client network with SV prices and consumer investments. We first describe a single-period two-stage Stackelberg pricing game between an SV and its consumers. We then state the mathematical results related to the pricing game equilibria for the monopoly setting under both, the heterogenous and the uniform pricing scenarios, and provide the relevant practical implications of the results. Finally, we address the problem of designing an optimal pricing mechanism for the binary pricing scenario, which is a special case of the heterogenous pricing scenario, but is NP-hard. For simplicity, we do not analyse the oligopoly setting of SVs for closed form expressions of game parameters at equilibria. However, we conduct simulation experiments on the oligoply setting (see Section 4). 
\subsection{Defining the Pricing Game}
Our proposed SV pricing mechanism entails a one-period, two-stage Stackelberg pricing game consisting of the following two steps. 
\begin{enumerate}
\item The SV chooses a price vector $\overrightarrow{p}$ so as to maximize its profits via the following optimization problem.
\[max_{\overrightarrow{p}}\sum_{i} p_{i}x_{i} - cx_{i},\]

We consider three types of consumer pricing scenarios in the paper:
\begin{itemize}
 \item \emph{\textbf{Scenario 1}} -  the SV does not price discriminate amongst its consumers and all elements of $\overrightarrow{p}$ are identical, i.e., $p_{i} = p, \,\forall i$. 
\item \emph{\textbf{Scenario 2}} (binary pricing) - the SV charges two types of prices per unit of user investment: a regular price denoted as $p_{reg}$ for each user in a particular category, and a discounted price, denoted as $p_{dsc}$ for other users,
\item \emph{\textbf{Scenario 3}} - the SV charges different prices to different consumers. 
\end{itemize}
\item Consumer $i$ chooses to consume $x_{i}$ units of self-defense products, so as to maximize his utility $u_{i}(x_{i}, ,\overrightarrow{x_{-i}}, p_{i})$ given the prices chosen by the SV. 
\end{enumerate}
Since the pricing strategy is the output of a dynamic consisting of two stages, we will analyze the subgame perfect Nash equilibria (SPNE) of this game, instead of just focussing on traditional Nash equilibria.  A strategy profile is a SPNE if it represents a Nash equilibrium of every subgame of the original game. Informally, this means that if (1) the players played any smaller game that consisted of only one part of the larger game and (2) their behavior represents a Nash equilibrium of that smaller game, then their behavior is a subgame perfect equilibrium of the larger game. Every finite extensive game, like ours, has a subgame perfect equilibrium \cite{mwg}. 

\subsection{Results - Pricing Strategy} \label{sec-results}
In this section, we state results related to the pricing strategy arising from the game equilibria and analyze the intuition and practical implications behind those results. We first comment on the equilibria of the second stage of the two-stage pricing game, \emph{given} a vector of prices $\overrightarrow{p}$. Given $\overrightarrow{p}$, the second stage of our pricing game is a subgame and we denote it as $G^{sub}$. We then have the following theorem. The proof of the theorem is in Section 6. 
\begin{theorem} 
$G^{sub}$ has a unique Nash equilibrium and is represented in closed form as 
\begin{equation}
x_{i} = BR(\overrightarrow{x_{-i}}) = \frac{\alpha_{i} - p_{i}}{2\beta_{i}} + \frac{1}{2\beta_{i}}\sum_{j\epsilon N}h_{ij}x_{j},
\end{equation}
where $BR(\overrightarrow{x_{-i}})$ is the strategic best response of user $i$ when other users in the network consume $\overrightarrow{x_{-i}}$. In the case when SV does not price discriminate its consumers, the Nash equilibrium vector of user investments is given by
\begin{equation}
\overrightarrow{x} = (Q - G)^{-1}(\overrightarrow{\alpha} - \overrightarrow{p}\overrightarrow{1}),
\end{equation}
where $\overrightarrow{p}$ is the optimal uniform per unit investment price charged
by the SV to all its consumers, and $Q$ is a matrix that takes values $2\beta_{i}$ at location $(i,j)$ if $i = j$ and zero otherwise.
\end{theorem}
\emph{Theorem Intuition and Implications:} The intuition behind a unique sub-game perfect Nash equilibrium is the fact that increasing one's consumption (security investment) incurs a positive externality on his peers, which further implies that the game involves \emph{strategic complementarities}\footnote{In economics and game theory, the decisions of two or more players are called strategic complements if they mutually reinforce one another.} \cite{mwg}, and therefore the equilibria are ordered. This monotonic ordering results in a unique NE \cite{topkis}. 
The benefit of dealing with a single equilibrium vector for $G^{sub}$ is the ease with which the SV can decide on its optimal strategy.

\textbf{Optimal Pricing Strategy} We now discuss the optimal pricing strategy for the SV given that the users self-protect according to the Nash equilibrium of $G^{sub}$. Before going into the details we first define the concept of a Bonacich centrality in a network of heterogenous users. The Bonacich centrality measure \cite{bonacich} is a sociological graph-theoretic measure of network influence. It assigns relative influence scores to all nodes in the network based on the concept that connections to high-scoring nodes contribute more to the score of the node in question than equal connections to low-scoring nodes. In our work, the Bonacich measure of a user reflects his influence on other users of the network via the positive externalities generated by him through his self-defense investments\footnote{The use of the concept of Bonacich centrality in externality driven networks is explained in \cite{ballester2006s}}. Formally, let $G$ be a matrix defining the logical network of $N$ users (consumers), and having in its entries the $h_{ij}$ values. Let $D$ be a diagonal matrix, and $\overrightarrow{w}$ be a weight vector. The weighted Bonacich centrality vector is given by
\begin{equation}
B(G, D, \overrightarrow{w}) = (I - GD)^{-1}\overrightarrow{w},
\end{equation}
where $(I - GD)^{-1}$ is well-defined and non-negative. 

We now have our first result regarding the optimal prices charged by the SV to its consumers. The proof of the theorem is in Section 6.
\begin{theorem}
The unique optimal price vector $\overrightarrow{p}$ charged by the SV is given by
{\footnotesize
\begin{equation}
\overrightarrow{p} = \frac{\overrightarrow{\alpha} + c\cdot\overrightarrow{1}}{2} + GQ^{-1}B(G', Q^{-1}, \overrightarrow{w'}) - G^{T}Q^{-1}B(G', Q^{-1}, \overrightarrow{w'}),
\end{equation}}
where $G' = \frac{G + G^{T}}{2}$ and $\overrightarrow{w'} = \frac{\overrightarrow{\alpha} - c\cdot\overrightarrow{1}}{2}$. 

In the case when the SV does not price discriminate its consumers, the unique optimal price (same for every consumer) charged per consumer is given by  
\begin{equation}
p = \frac{1}{2}\frac{\overrightarrow{1}^{T}(Q - G)^{-1}(\overrightarrow{\alpha} + c\overrightarrow{1})}{\overrightarrow{1}^{T}(Q - G)^{-1}\overrightarrow{1}}.
\end{equation}
\end{theorem}
\emph{Theorem Intuition and Implications:} The optimal price vector in the no price discrimination case is independent of individual node centralities, whereas in the price discrimination case the optimal price vector depends on the Bonacich centrality of individual users. The intuition behind the result is the fact that users tend to invest in security mechanisms proportional to their Bonacich centrality (and in turn generate proportional amount of network externalities) in the Nash Equilibrium \cite{palhui1}\cite{palhui}. Therefore it makes sense for the SV to charge users based on their Bonacich centralities when price discrimination is possible. In addition, Equation (5) has three parts to its price vector expression. The first subexpression is that part of the price vector that is topology invariant. The second subexpression is that part of the price vector that is topology variant and reflects the markup price charged to network users for the positive externalities they gain due to the security investments of other users. Finally, the third subexpression is that part of the price vector which is topology variant and reflects user price discounts for the benefits they provide to other users in the network through their security investments. 

We now state the following result regarding profit amounts made by an SV in pricing scenarios 1 and 3. The proof of the theorem is in Section 6. 
\begin{theorem}
The profits, $P_{0}$ and $P_{1}$,  made by an SV when the latter does not (does) account for user investment externalities are given by 
does not: 
\begin{equation}
P_{0} = \left\{\left(\frac{\overrightarrow{\alpha} - c\cdot\overrightarrow{1}}{2}\right)^{T}(Q - G)^{-1} \left(\frac{\overrightarrow{\alpha} - c\cdot\overrightarrow{1}}{2}\right)\right\}
\end{equation}
and
does: 
\begin{equation}
P_{1} = \left\{\left(\frac{\overrightarrow{\alpha} - c\cdot\overrightarrow{1}}{2}\right)^{T}(Q - G')^{-1} \left(\frac{\overrightarrow{\alpha} - c\cdot\overrightarrow{1}}{2}\right)\right\}.
\end{equation}
Assuming $Q - G$ to be positive definite, the bounds of the ratio of profits in these two cases is given by 
\begin{equation}
0 \le\frac{1}{2} + \lambda_{min} (K) \le \frac{P_{0}}{P_{1}} \le \frac{1}{2} + \lambda_{max}(K)\le 1,
\end{equation}
where $R = Q - G$ and $\lambda_{min}(\cdot)$, $\lambda_{max}(\cdot)$ denote the minimum and maximum eigenvalues of their arguments respectively, and $K = \left(\frac{RR^{-T} + R^{T}R^{-1}}{4}\right)$.
\end{theorem}
In brief, the main intuition behind the theorem result is that an SV can make more profits when accounting for network externalities compared to when it does not. The inuition here is that more informed pricing implies more profit. We postpone a more detailed explanation of the theorem intuitions and its implications to Section 4, as we find it appropriate to first plot the ratio of $\frac{P_{0}}{P_{1}}$ for instances of different graph topologies to gain a better understanding of the theorem implications.  

\subsection{The Case of Binary Pricing}
In reality, the idea of charging multiple different prices to various consumers may not be very practical to implement, primarily because of the difficulty to select different classes of users paying different prices. To make things simpler, an SV can opt to charge two types of prices for two different classes of consumers: (i) a discounted price, $p_{dsc}$, for consumers who have significant positive influence on the security of a network based on their network location and the amount of investments made, (ii) and a regular price, $p_{reg}$ for the other consumers. Thus, the first goal of an SV is to determine the subset of consumers who should be offered the discounted price so as to maximize the SVs own profits. 

Given that $p_{reg}$ and $p_{dsc}$ are exogenously specified, the profit optimization problem for an SV is given by 
\[Maximize\,\, (\overrightarrow{p} - c\overrightarrow{1})^{T}(Q - G)^{-1}(\overrightarrow{\alpha} - \overrightarrow{p})\]
\[s.t.\,\, p_{i}\,\epsilon\,\{p_{reg}, p_{dsc}\},\,\forall i\,\epsilon\,N.\]
Note here that the expression, $(Q - G)^{-1}(\overrightarrow{\alpha} - \overrightarrow{p})$, in the objective function is the NE investment amount of users in self-defense mechanisms. Thus, we have a combinatorial optimization problem for maximizing the profits of an SV. In order to investigate the tractability of the problem, we first formulate it in the following manner: 
\[OPT: \,\,Maximize\,\, (\delta\overrightarrow{y} + c'\overrightarrow{1})^{T}(Q - G)^{-1}(\overrightarrow{\alpha'} - \delta\overrightarrow{y})\]
\[s.t.\,\, y_{i}\,\epsilon\,\{-1,1\},\,\forall i\,\epsilon\,N.\]
Here $\delta = p_{reg} - p_{T}$, where $p_{T} = \frac{p_{reg} + p_{dsc}}{2}$, $\overrightarrow{a'} = \overrightarrow{a} - p_{T}$, and $c' = p_{T} - c \ge \delta$. Note that using these variables, the feasible price allocation can be expressed as $\overrightarrow{p} = \delta\overrightarrow{y} + p_{T}$. Our next result comments on the intractability of solving OPT. The proof of the result which is in Section 6, is based on the reduction of OPT from the well-known MAX-CUT problem \cite{gjr}. 
\begin{theorem}
Given that $p_{reg}$ and $p_{dsc}$ are exogenously specified by the SV, in the binary pricing case an SV's profit optimization problem, OPT, is NP-hard.
\end{theorem}
\emph{Theorem Intuitions and Implications:}  The $i,j$th entry of $(Q - G)^{-1}$ gives a measure of how much the edge between $i$ and $j$ contributes to the centrality of consumer $i$. Therefore, the MAX-CUT interpretation roughly suggests that the optimal solution of the pricing problem is achieved when the monopolist tries to price discriminate consumers that influence each other significantly, however, at the same time takes into account the consumers' value of consumption in the absence of network effects. The NP-hard nature of the pricing problem implies the impracticality of computing optimal prices in practice and thus drives the need to design schemes to computing optimal binary prices up to a certain acceptable approximation. The following theorem states the result of approximating optimal prices charged in the binary pricing case. The theorem exploits the relation of OPT to the MAX-CUT problem, and establishes that there exists an algorithm that provides a solution with a provable approximation guarantee. The proof of the theorem is in Section 6. 
\begin{theorem}
Let $W_{OPT}$  be the optimal value for problem OPT. Then, there exists a randomized polynomial-time algorithm that outputs a solution with objective value $W_{alg}$ such that $E[W_{alg}] + r > 0.878(W_{OPT} + r)$, where
{\small
\[ r = \delta^{2}\overrightarrow{1}^{T}A\overrightarrow{1} + \delta\overrightarrow{1}^{T}|A\overrightarrow{a'} - A^{T}c'\overrightarrow{1}| - c'\overrightarrow{1}^{T}A\overrightarrow{a'} - 2\delta^{2}Trace(A),\]}
and 
\[A = (Q - G)^{-1}\]
\end{theorem}
\emph{Theorem Implications:} Clearly, if $r \le 0$, which, is the case when $\delta$ is small, this algorithm provides at least an 0.878-optimal solution of the problem. On the other hand, if $r > 0$, we obtain 0.878 optimality after a constant $(r)$ addition to the objective function. This suggests that for small $r > 0$, the algorithm still provides near-optimal solutions.

\section{Numerical Evaluation}
In this section, we perform an extensive numerical evaluation study of monopoly and oligopoly SV pricing markets comprising of (i) autonomous SVs and (ii) rational, non-colluding consumers connected through a logical communication network. Through numerical simulations, we aim to address the following questions:
\begin{enumerate}
\item What is the relationship between SV prices and the client network topology? 
\item For any given pricing scenario, how does the network location of a consumer relate with their total security investment amount?
\item Are the SV pricing schemes fair to network users?
\item How does the peformance of SV pricing schemes differ between the monopolistic and oligopolistic market settings?
\end{enumerate}

\subsection{Parameters and Markets}
\noindent\textbf{Parameter Setting:} We deal with preferential attachment (PA) and random tree network topologies in this work (see Section 4.2 for details). For both monopolistic and oligopolistic market settings, the model parameters (e.g., PA and random tree networks) are as follows: $c = 0.5, \alpha_{i} = 2$, $\beta_{i} \in \{2, 2.5, 3\}$ (for the preferential attachment mechanism), and $\beta_{i} = \frac{|G|}{20}$ (for random trees), for all $i\,\epsilon\,N$. The parameters $\alpha_{i}$ and $\beta_{i}$ are chosen to make sure that the utility function of each client is twice differentiable and strictly increasing. We assume that the influence matrix $G$ is such that for all $i$, $h_{ij}$ = $\frac{1}{d_{i}}$, where $d_{i}$  is the number of non-negative entries in row $i$ of $G$.  Due to several competing SVs serving clients on the same network $G$ in an oligopolistic setting, we assume that information on the estimates of externalities generated by consumer investments can be exchanged for free amongst the competing SVs in a truthful manner. 

\noindent \textbf{Market Setting:} For the purposes of simulation, we assume a monopolistic market setting with \emph{one} SV, and an oligoplistic setting with \emph{two} SVs. To simulate the oligopolistic competition between the SVs, each SV first plays a Stackelberg price game amongst its clients as in the monopoly setting to reach a Stackelberg equilibrium price vector that is either single priced or binary priced, and then repeatedly plays with its competing SV in multiple rounds to converge upon a sub-game perfect market Nash equilibrium price vector. This dynamic repeated pricing game between competing SVs is played with the mindset of an SV being able to prevent its clients from switching to a different SV after each round. Finally, on convergence, each SV and its clients play the sub-game perfect Nash equilibrium strategies. In our work we practically obtain the Stackelberg equilibrium (for the monopoly setting) and the market Nash equilibrium (for the oligopoly setting) price vectors in an \emph{iterative} fashion using the standard \emph{fictitous play} technique in game theory \cite{ft}. 

\subsection{Topology Formation}
We choose preferential attachment graphs as one topology type as they represent real world social/logical network interactions \cite{abr1}. A random graph formed by the PA process can have two extremes: (i) a new born user can influence users born earlier, i.e., $G^{1}_{ij} > 0$ for all $i,j$  and $j$ born after $i$, and (ii) only older users influence new users, i.e., $G^{1}_{ij} > 0$ for all $i,j$ and $j$ born before $i$. We can thus form a family of PA graphs parameterized by $\mu\,\epsilon\,[0,1]$, which we call the `influence value', such that $G^{\mu}$ is a linear combination of $G^{1}$ and $G^{2}$, i.e., $G^{\mu} = \mu G^{1} + (1 - \mu)G^{2}$. Here $\mu$ is a parameter that controls the influencing nature of a user in a PA graph w.r.t. the positive externality effects of their security investments made on users born before and later. A $\mu$ value of 1 generates random graph $G^{1}$, whereas a $\mu$ value of 0 generates random graph $G^{2}$. For networks of size 500, we generate 50 random PA graphs for each different value of $\mu$ ranging from 0 to 1 in steps of 0.1. As an example, each point in a sub-plot in Figure 1 is the average of the 50 random $\frac{P_{0}}{P_{1}}$ values obtained per value of $\mu$ (the x-axis). Each sub-plot is the average of 50 graphs for a particular value of $\beta$, the scale-free exponent parameter for PA graphs, $\beta$ (in many papers also denoted as $\gamma$) is known to generally lie between $[2,3]$, and for our experiments we choose three discrete values of $\beta$: 2, 2.5, and 3. 

We choose tree topologies as they often represent corporate and enterprise social networks. \emph{Since a single SV is likely to serve these sort of networks, we do not consider and oligopolistic setting for these networks.} In this work, we consider random trees of specific types. Given a constant $\lambda > 0$, a depth-$d$ Poisson tree $T(\lambda, d)$ with parameter $\lambda$ where depth $d$ is constructed as follows: the root node has degree which is a random variable distributed according to a Poisson distribution with parameter $\lambda$. All the children of the root have outdegrees which are also random, with the same distribution. We continue this process until either the process stops at some depth $d' < d$, where no node in level $d'$ has any children or until we reach level $d$. In this case, all the children of nodes in level $d$ are deleted and the nodes in level $d$ becomes leaves. In this paper, we fix $d = 999999$, to represent $\d = \infty$. Note that star topologies are formed as a special case of tree topologies. We generate $\frac{P_{0}}{P_{1}}$ plots for tree graphs of size 500 in Figure 2, in the same manner as we generate PA graphs above, i.e., based on influence values. Each point in the sub-plots is the average of 50 instances for a fixed value of $\lambda$.  For our work we let $\lambda \in \{1,3,5\}$. For the purposes of simulation we assume for all $i$ that $\beta_{i} = \frac{|G|}{20}$ and $\alpha_{i} - c = 1$. We also assume that for each $i$ the $h_{ij}$ values are equal to $\frac{1}{d_{i}}$, where $d_{i}$ is the number of non-negative entries in row $i$ of $G$.
\begin{figure*}[htb]
\centering
 \begin{tabular}{@{}ccc@{}}
  \includegraphics[width=0.27\textwidth]{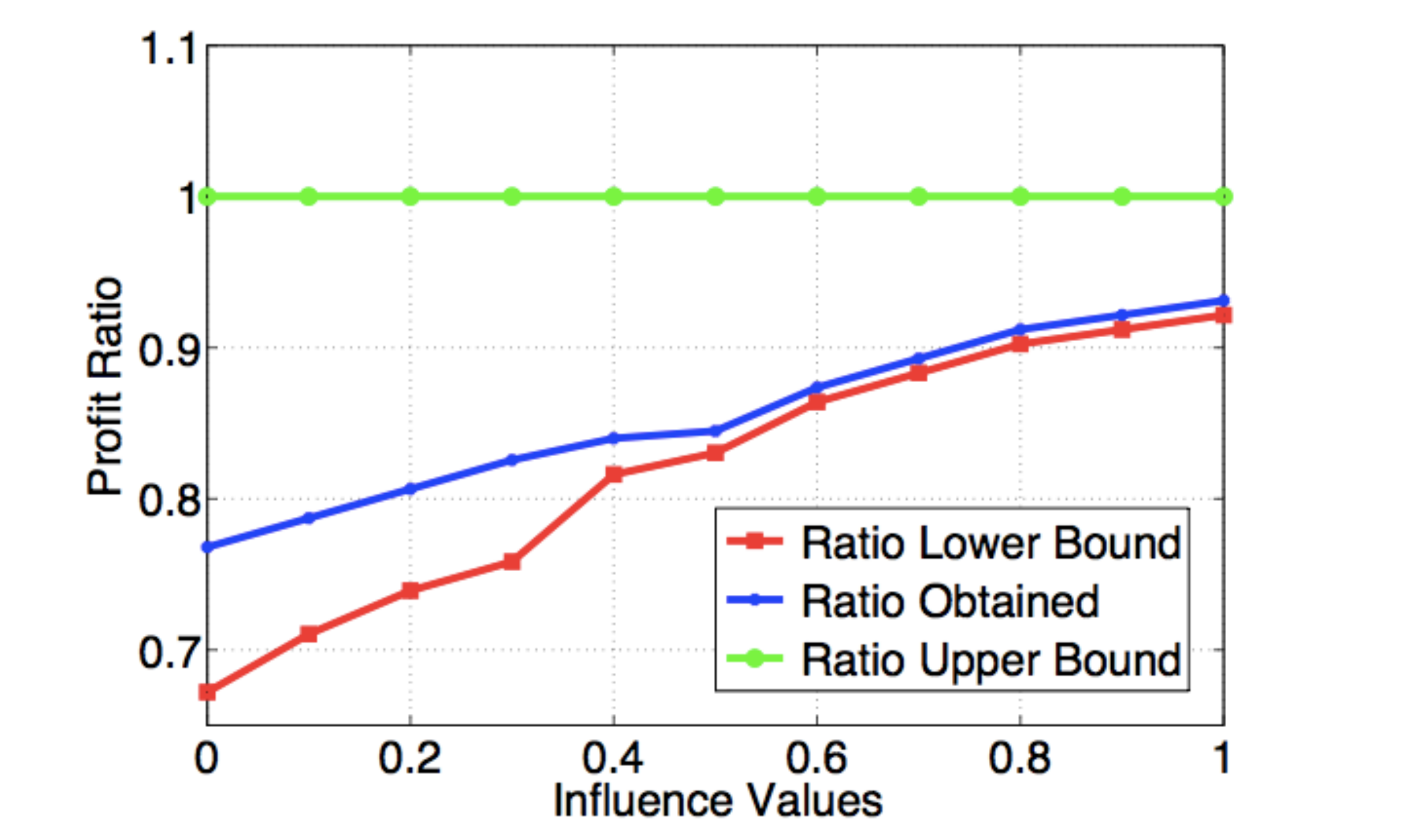} &
  \includegraphics[width=0.27\textwidth]{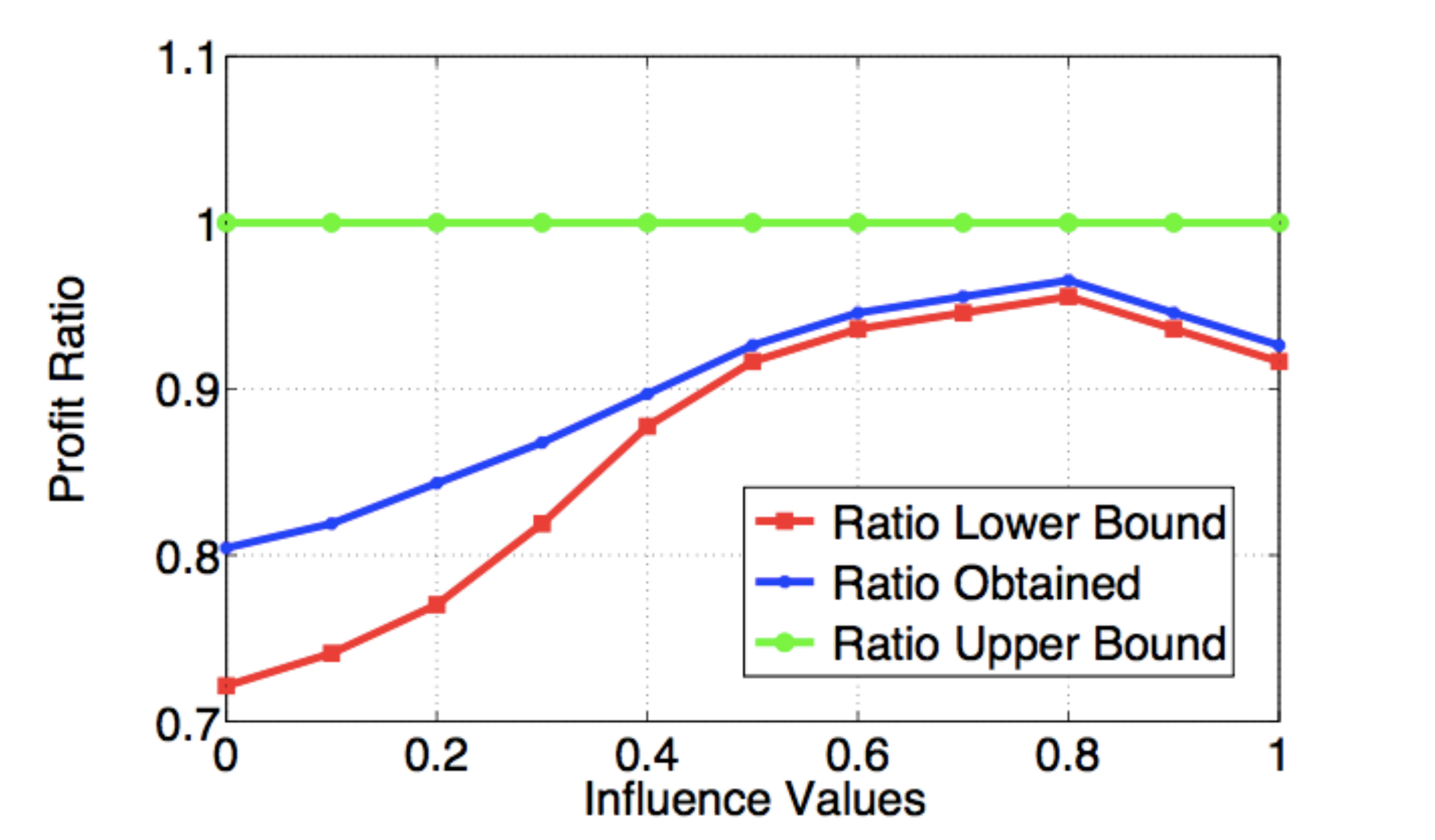} &
  \includegraphics[width=0.27\textwidth]{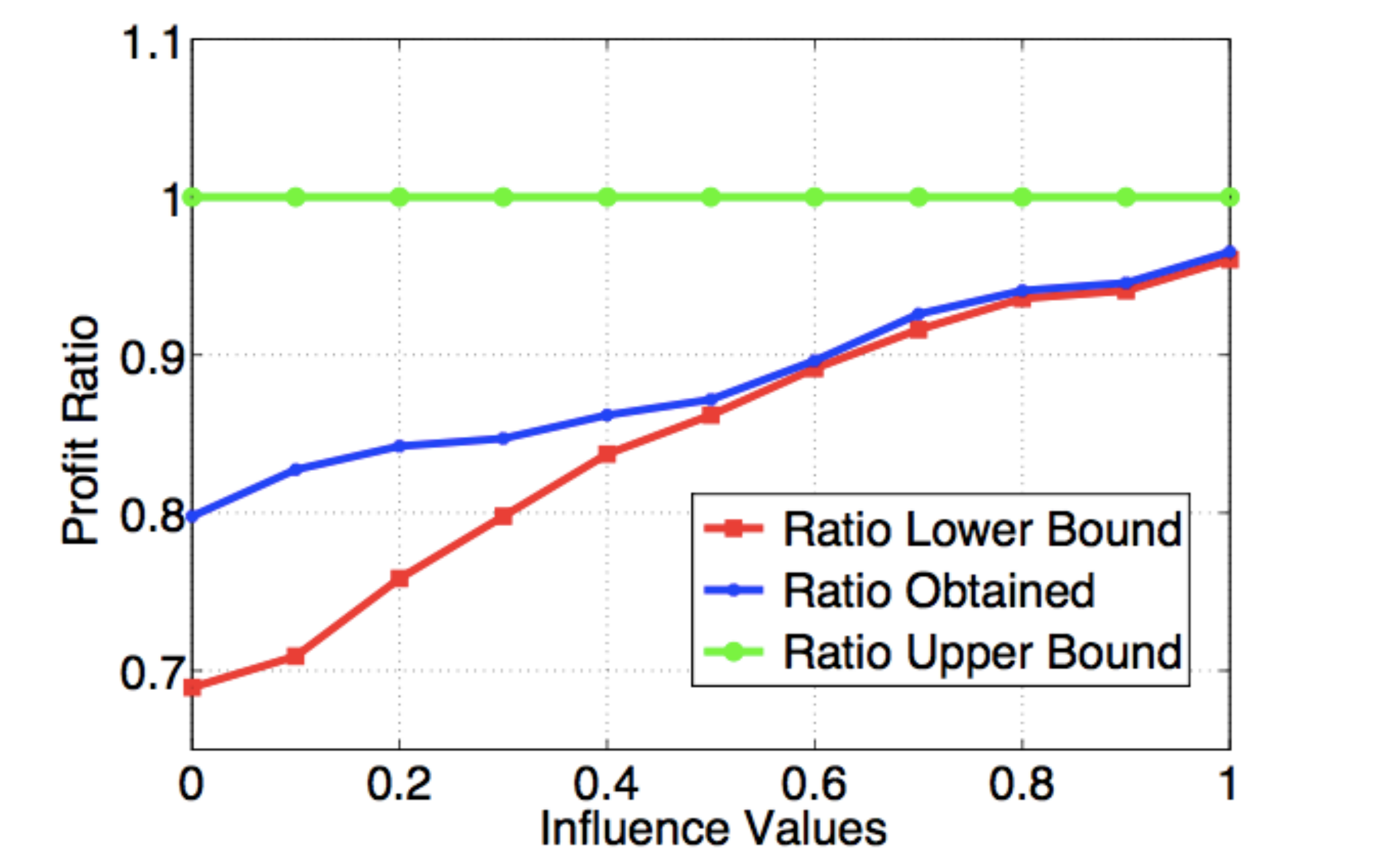}
\end{tabular}
 \caption{500 Node Profit Ratio (Monopoly) for (a) $\beta = 3$ (left), (b) $\beta = 2.5$ (middle), and (c) $\beta = 2$ (right) [PA Graphs]}
\end{figure*}

\begin{figure*}[htb]
\centering
 \begin{tabular}{@{}ccc@{}}
  \includegraphics[width=0.27\textwidth]{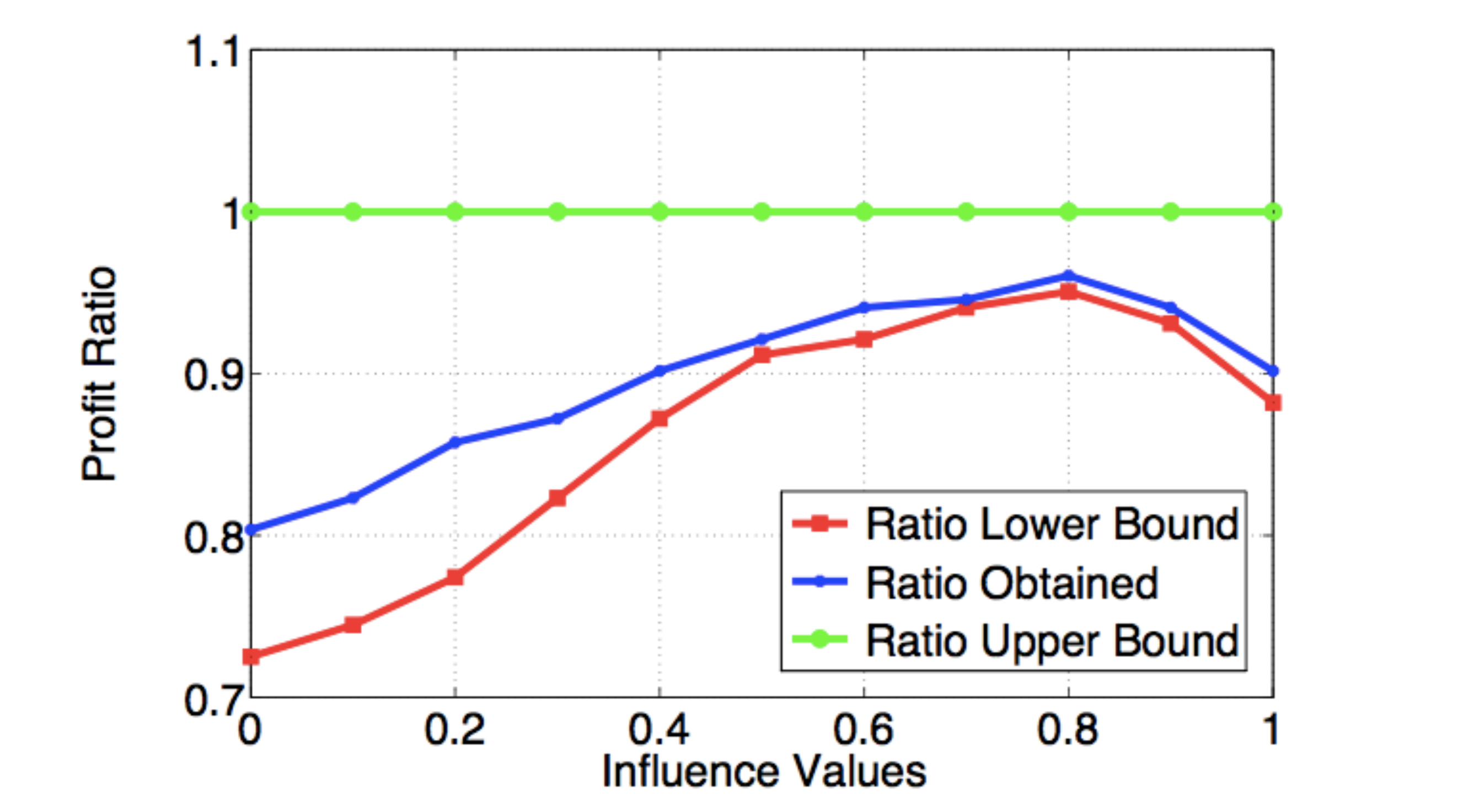} &
  \includegraphics[width=0.27\textwidth]{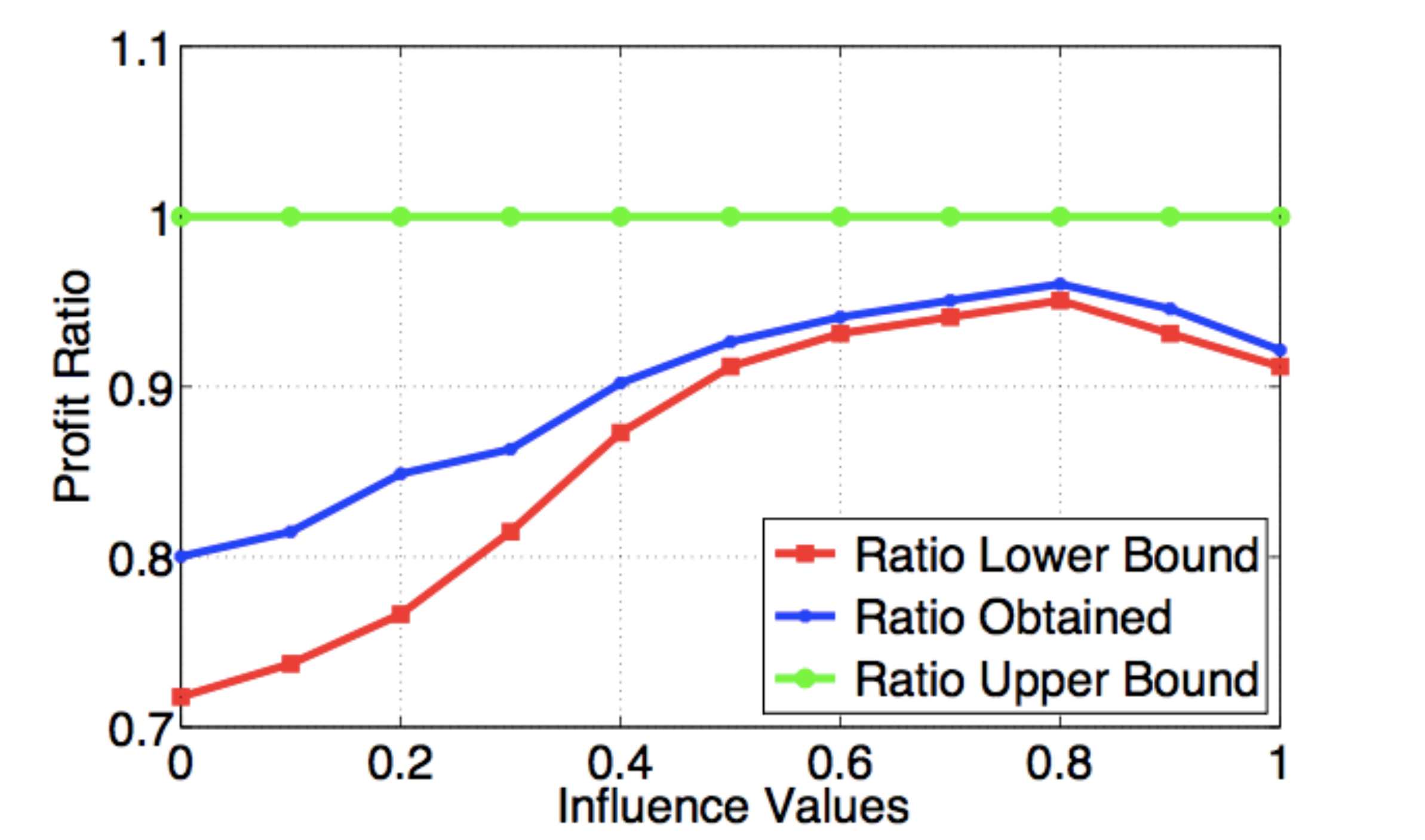} &
  \includegraphics[width=0.27\textwidth]{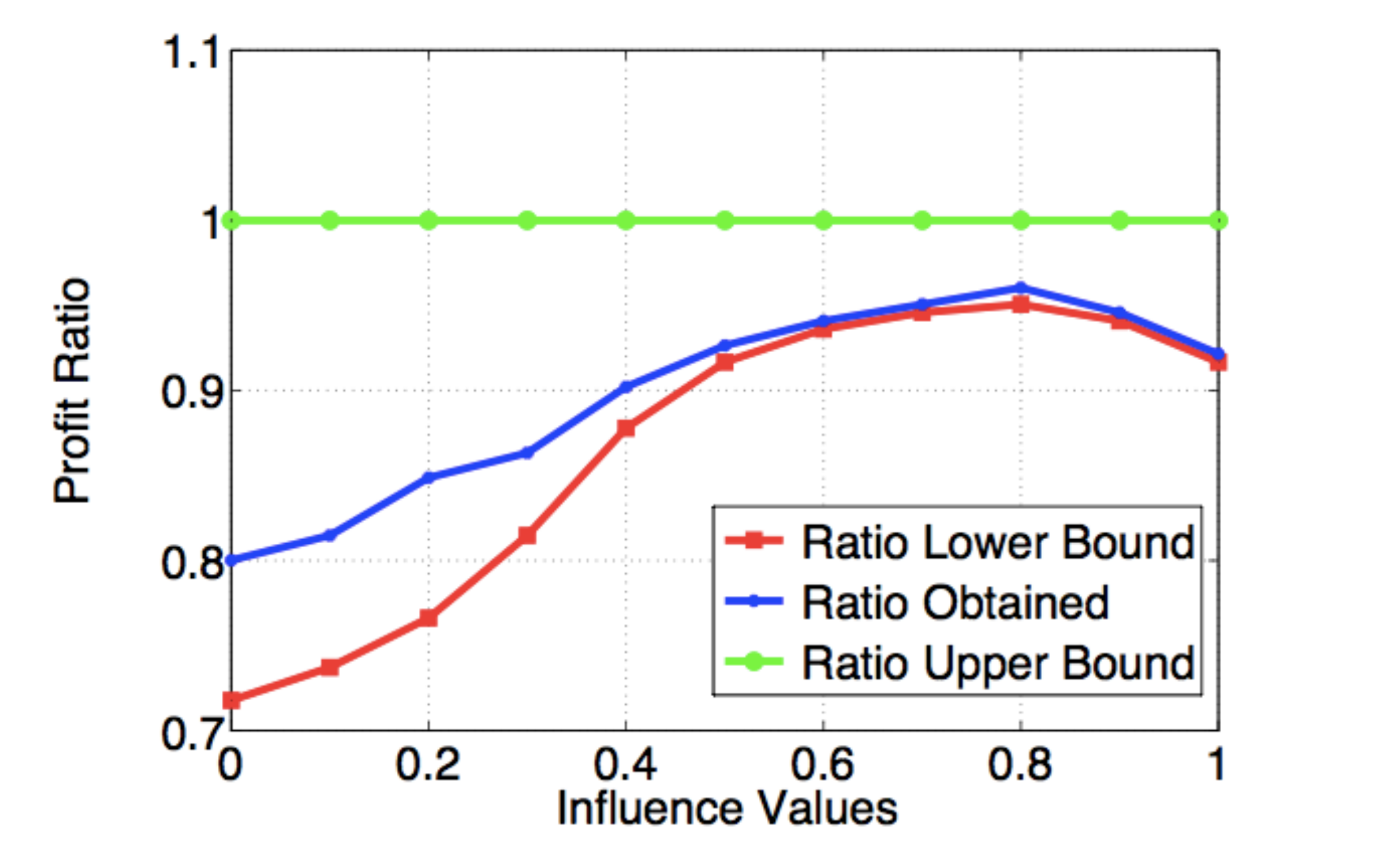}
\end{tabular}
 \caption{500 Node Profit Ratio (Monopoly) for (a) $\beta = 1$ (left), (b) $\beta = 3$ (middle), and (c) $\beta = 5$ (right) [Tree Topology]}
\end{figure*}

\begin{figure*}[htb]
\centering
 \begin{tabular}{@{}ccc@{}}
  \includegraphics[width=0.27\textwidth]{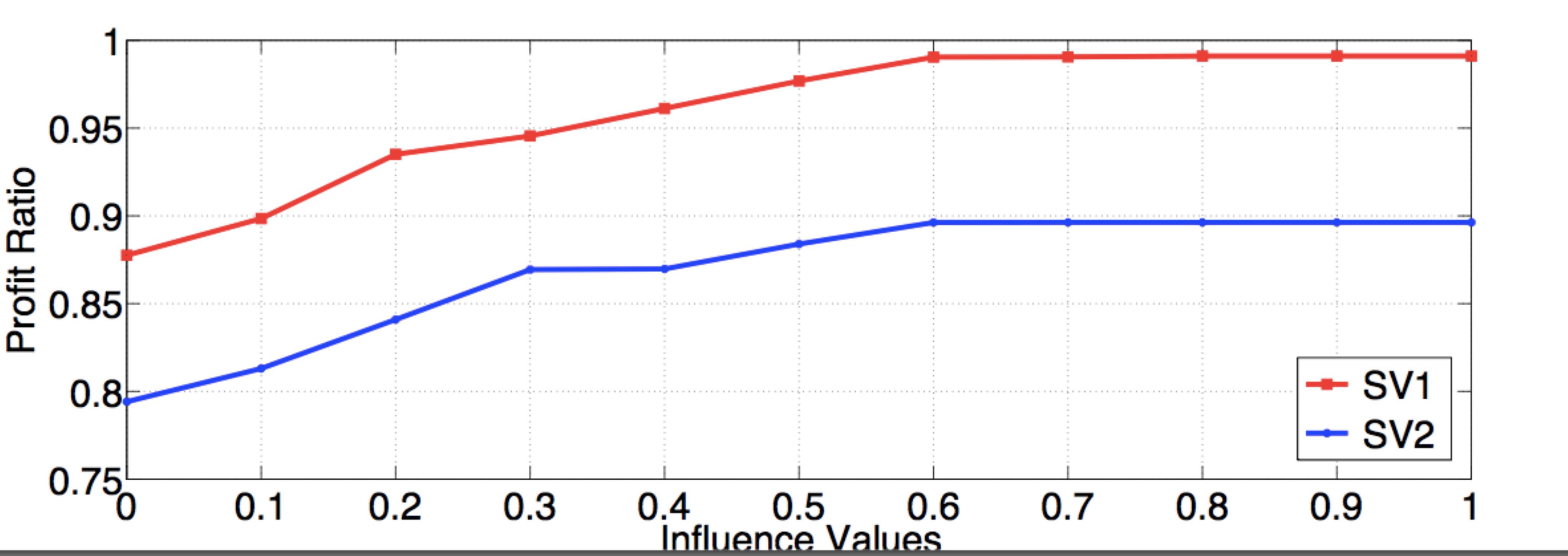} &
  \includegraphics[width=0.27\textwidth]{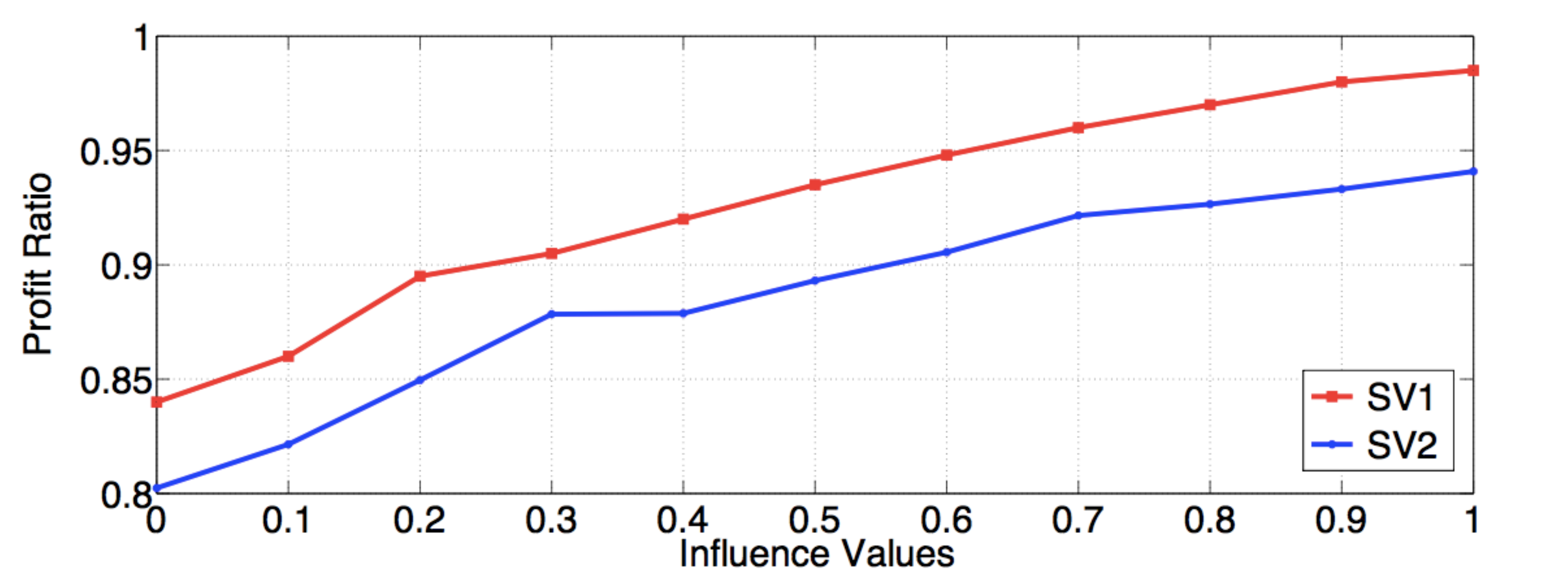} &
  \includegraphics[width=0.27\textwidth]{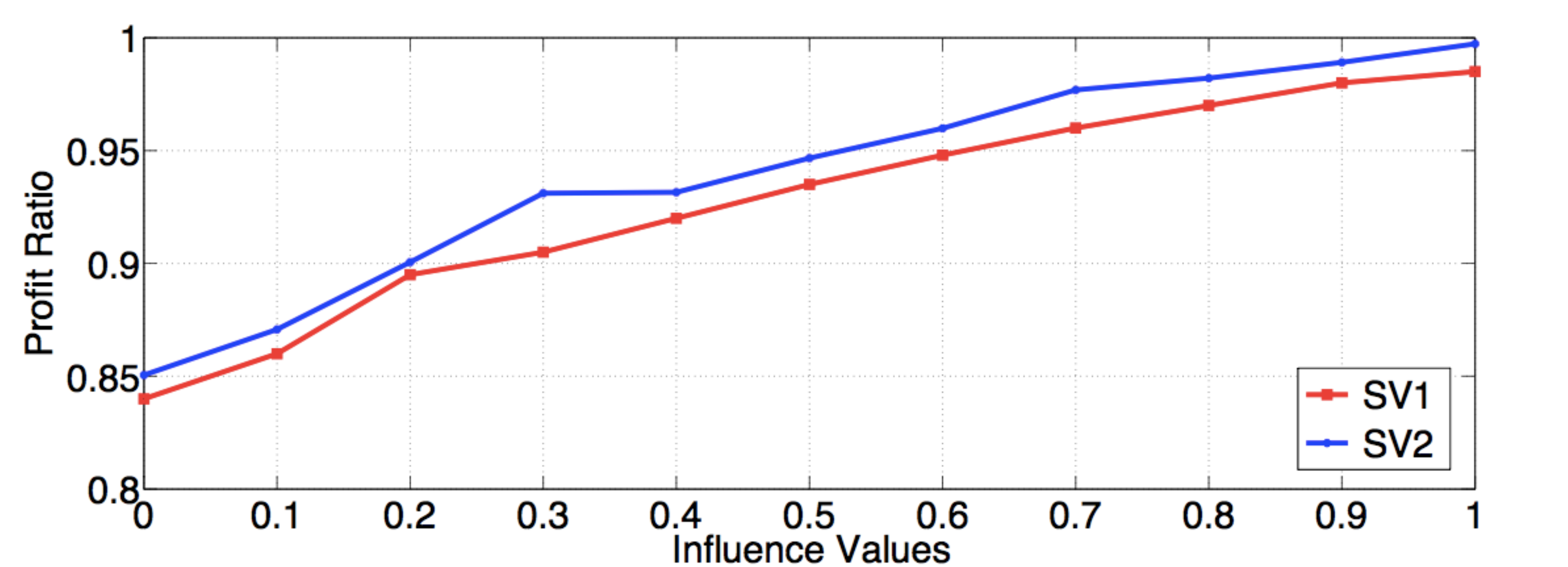}
\end{tabular}
 \caption{Profit Ratio (Oligopoly with 3 iterations) for (a) $\beta = 1$ (left), (b) $\beta = 3$ (middle), and (c) $\beta = 5$ (right) [PA Graphs]}
\end{figure*}

\begin{figure*}[htb]
\centering
 \begin{tabular}{@{}ccc@{}}
  \includegraphics[width=0.27\textwidth]{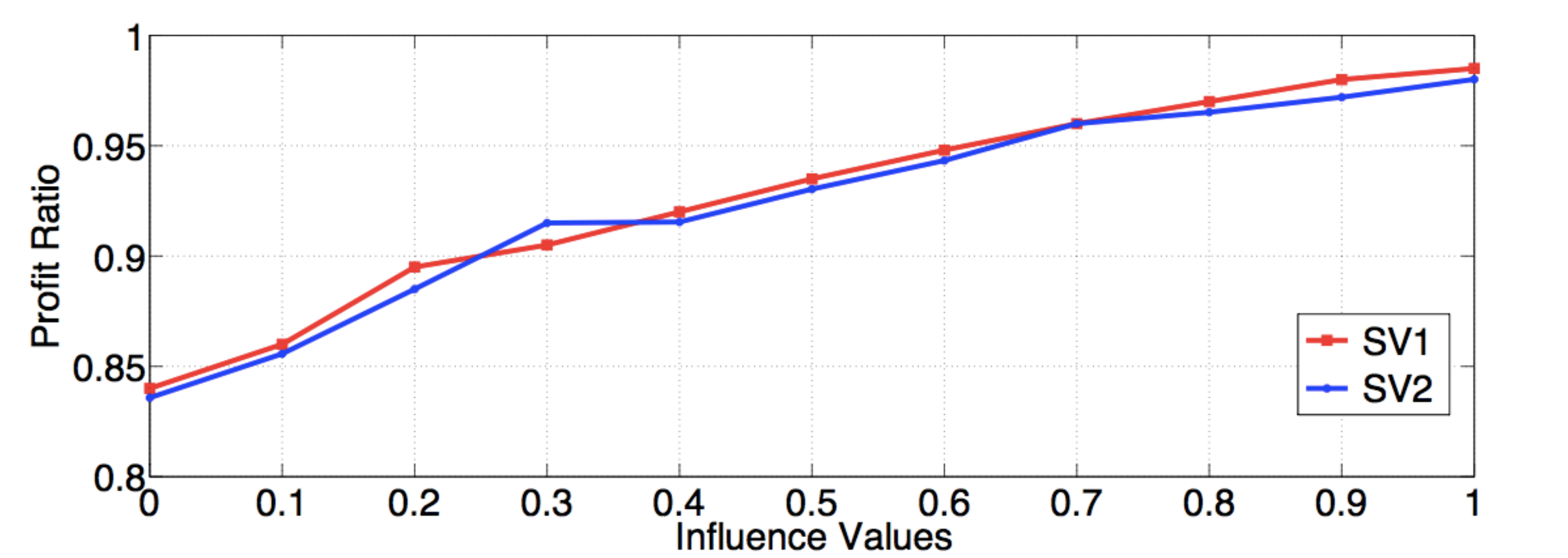} &
  \includegraphics[width=0.27\textwidth]{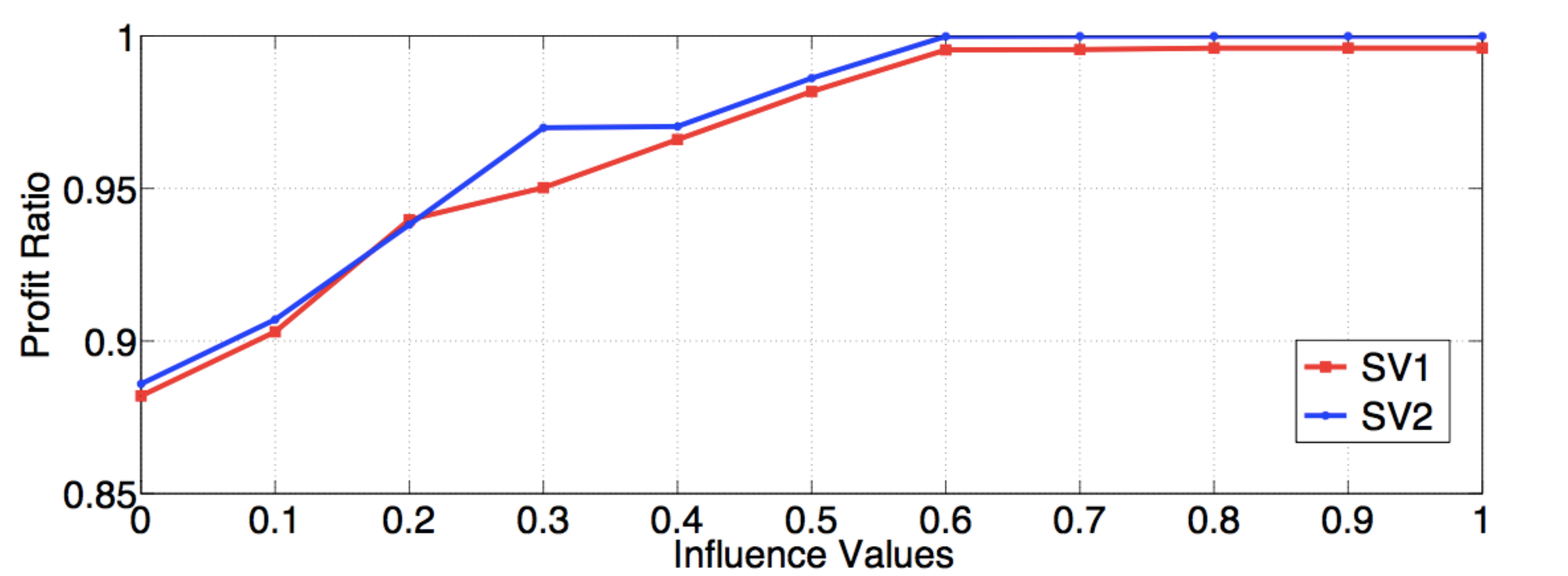} &
  \includegraphics[width=0.27\textwidth]{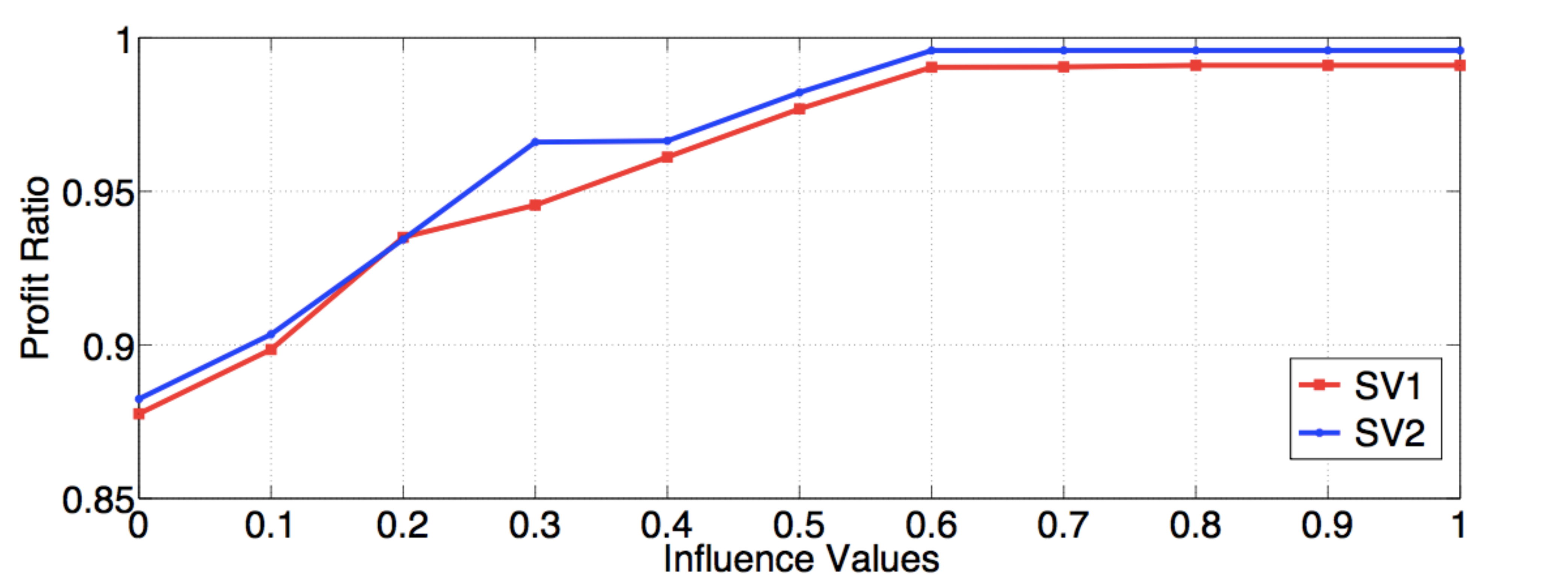}
\end{tabular}
 \caption{500 Node Profit Ratio (Oligopoly) for (a) $\beta = 1$ (left), (b) $\beta = 3$ (middle), and (c) $\beta = 5$ (right) [PA Graphs]}
\end{figure*}

\begin{figure*}[htb]
\centering
 \begin{tabular}{@{}cc@{}}
  \includegraphics[width=0.30\textwidth]{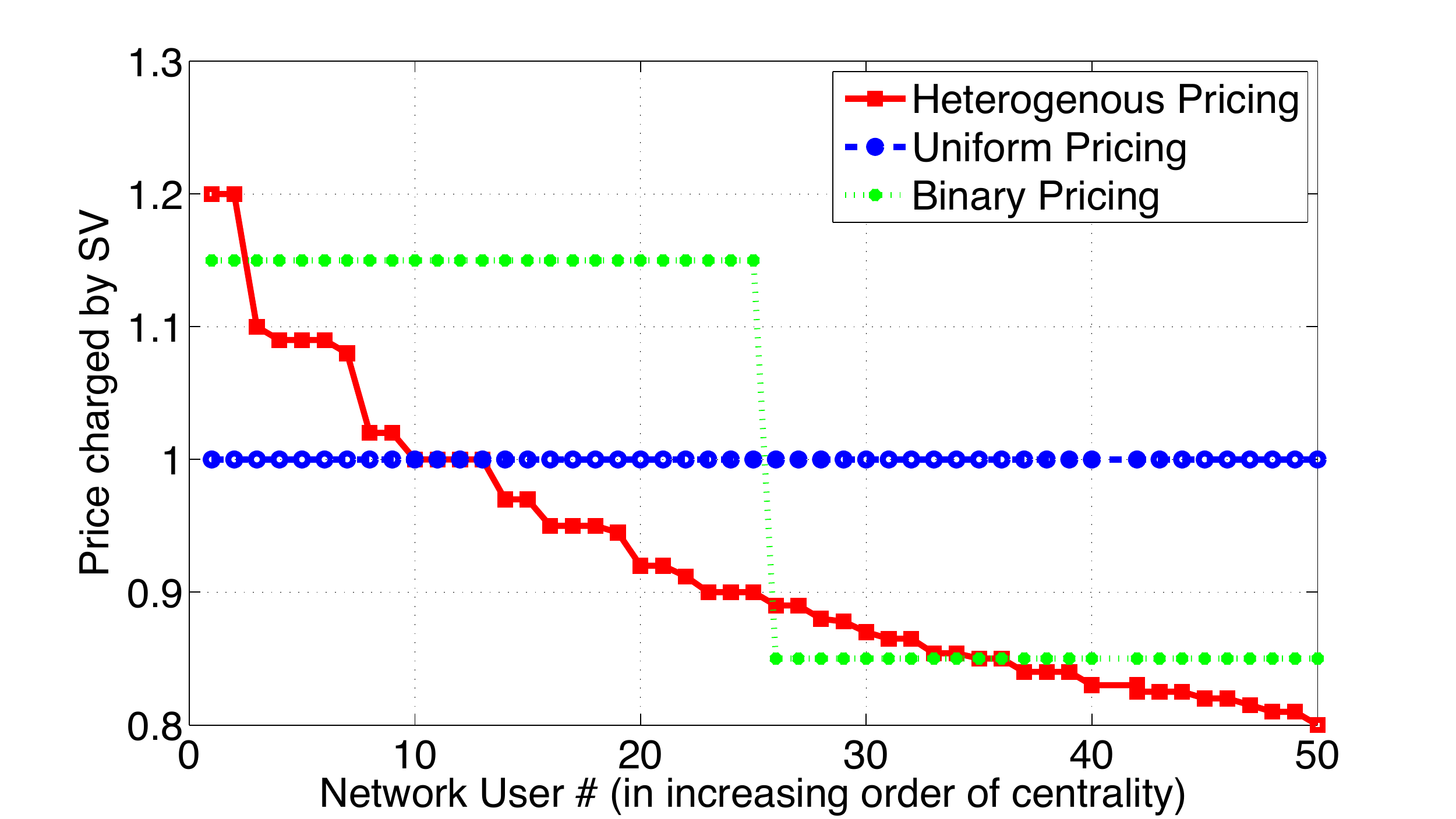} &
  \includegraphics[width=0.30\textwidth]{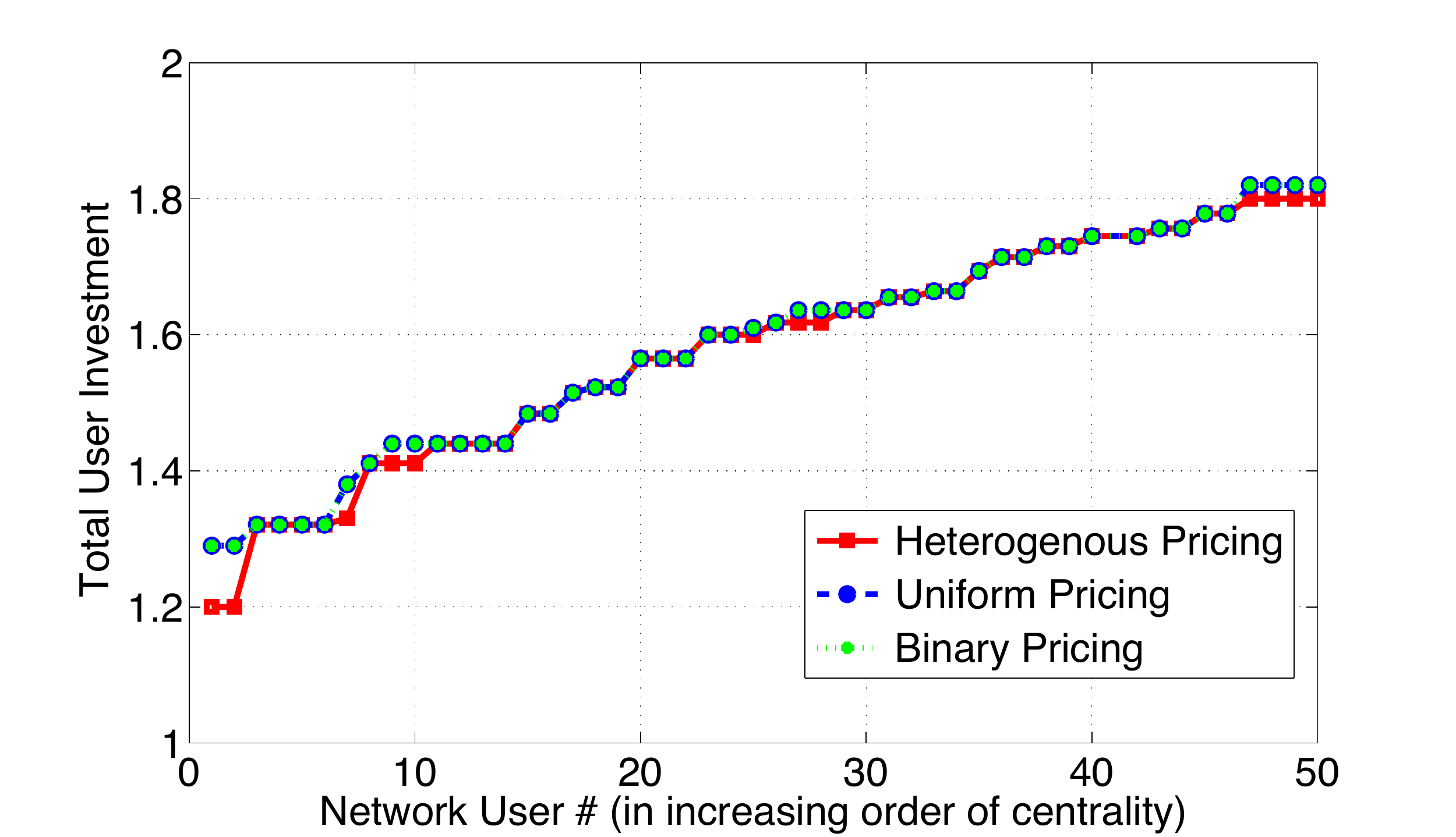} 
\end{tabular}
 \caption{50-Node Plots for (a) Per-Unit SV Prices (left) and (b) Total User Investment (right) when $\beta = 3$(PA Graph Topology)}
\end{figure*}

\begin{figure*}[htb]
\centering
 \begin{tabular}{@{}cc@{}}
  \includegraphics[width=0.30\textwidth]{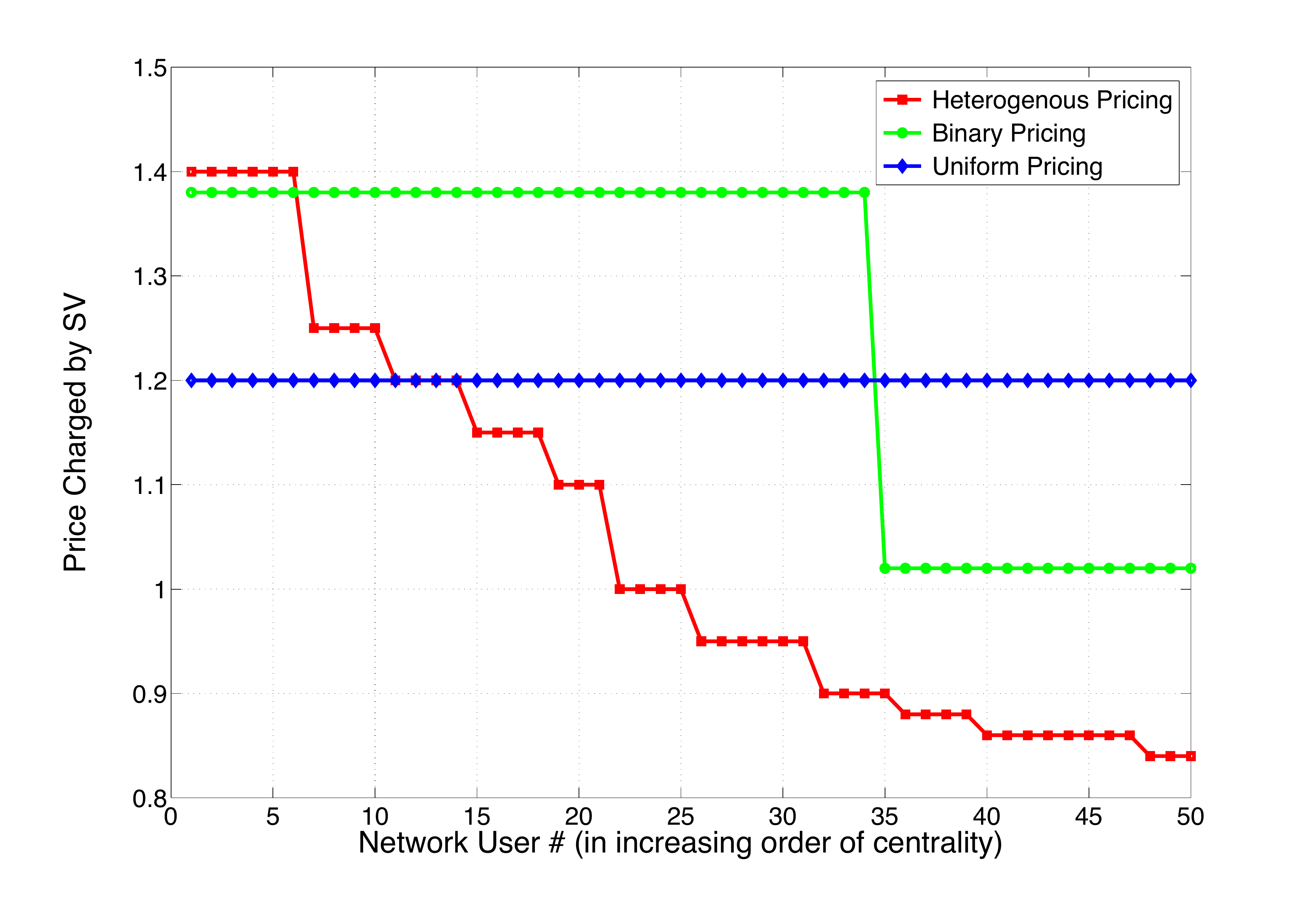} &
  \includegraphics[width=0.30\textwidth]{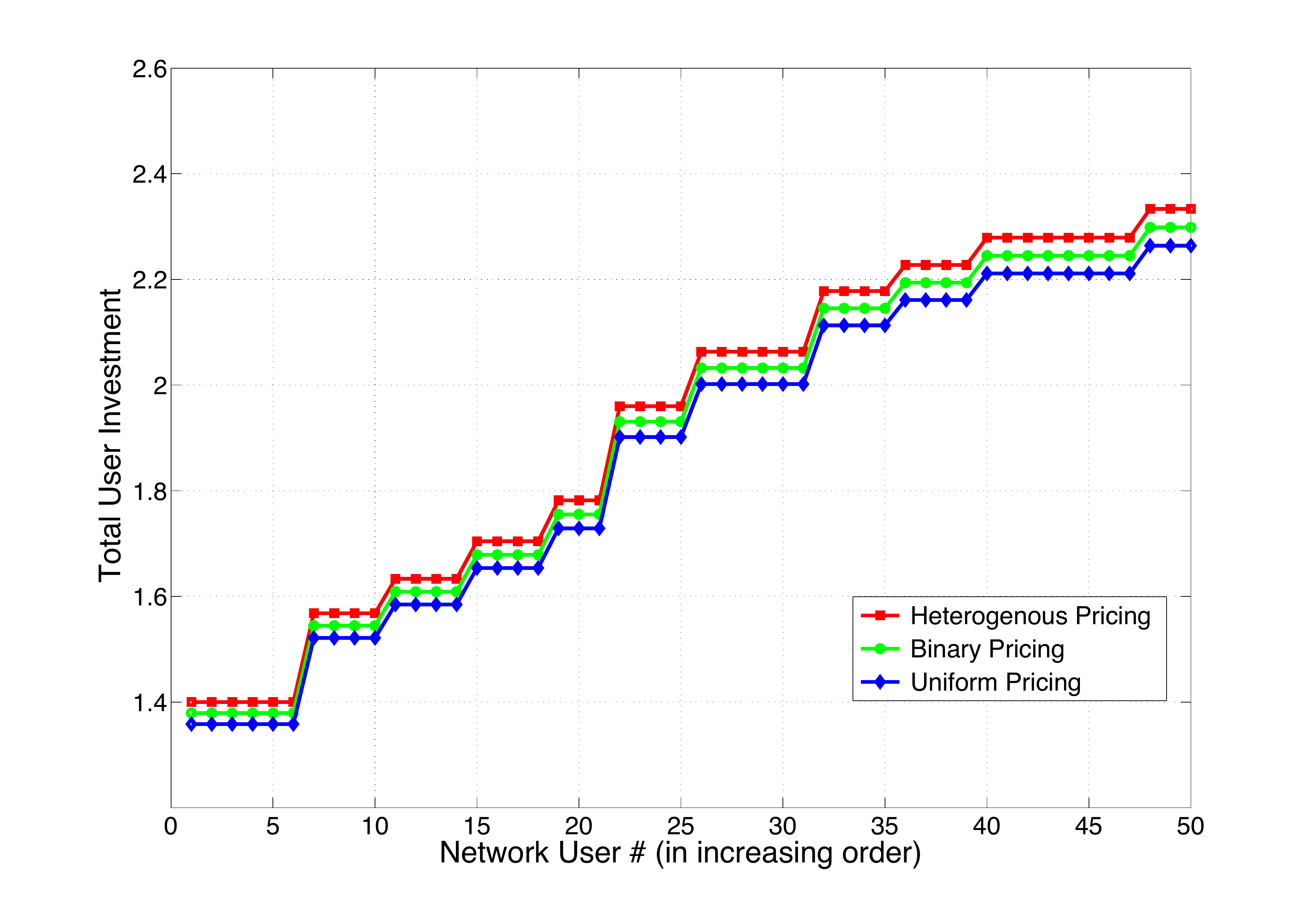} 
\end{tabular}
 \caption{50-Node Plots for (a) Per-Unit SV Prices (left) and (b) Total User Investment (right) when $\lambda = 3$ (Tree Topology)}
\end{figure*}

\begin{figure*}[htb]
\centering
 \begin{tabular}{@{}cc@{}}
  \includegraphics[width=0.30\textwidth]{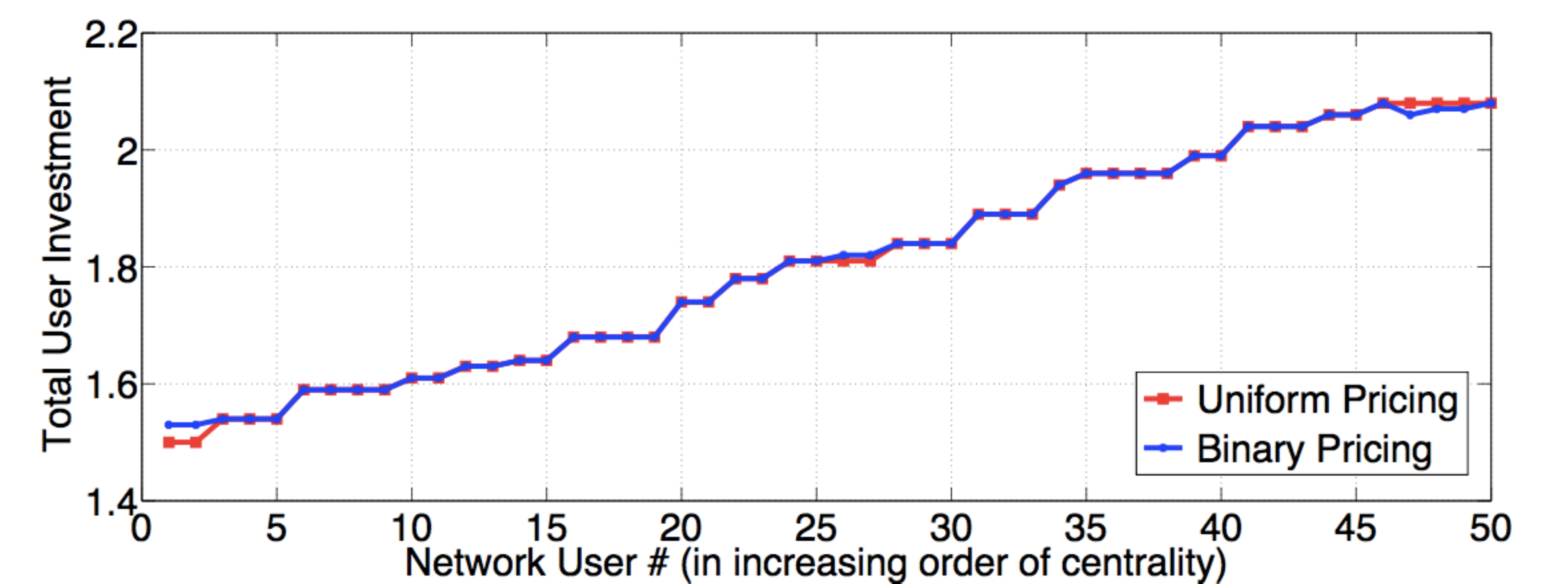} &
  \includegraphics[width=0.30\textwidth]{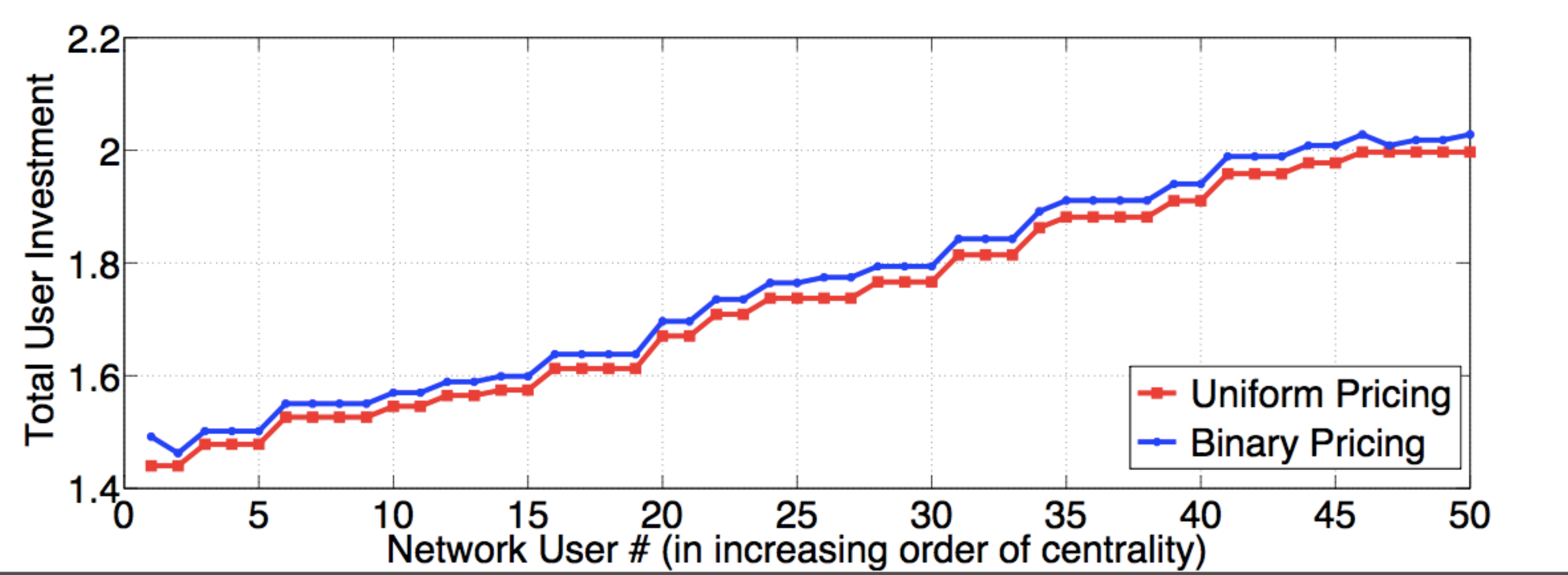} 
\end{tabular}
 \caption{50-Node Plots for Total User Investment when $\beta = 3$ (PA Graphs) in an Oligopoly Setting}
\end{figure*}

\subsection{Plot Observations and Insights}
\noindent \textbf{Profit Improvement:} Our plot results for(i) the preferential attachment graphs for both the \emph{monopoly and oligopoly setting} (see Figures 1,3) and (ii) tree topology graphs for the monopoly setting only (see Figure 2) show that (a) the provable \emph{profit ratio bounds are not reasonably tight enough, and are less than 1\footnote{Here `1' is the trivial upper bound. However, for the simulations in this work, the non-trivial upper bound obtained is very close to 1.}, } implying the fact that an SV can do better in terms of profit when it has full information compared to the case when it is not informed about consumer externality values and their network location properties, and (b) the gap between the ratio obtained from simulations and its lower bound decreases with increase in $\mu$ (influence) values. The reason for the trend in (b) can be explained as follows: note that the gap between the ratio obtained by simulations and its lower bound denotes the profit loss of the security vendor from ignoring the network externalities during the pricing process. This gap will be larger for lower values of $\mu$ as it implies that nodes with higher centrality values are affecting nodes with low centrality values and the insurer is not taking this externality into account while charging its clients, thereby reducing profits. When $\mu$ values are high, the insurer does not take into account the externality effects of low centrality users on high centrality users, which does not affect the profits as much.

\noindent \textbf{Implications from Theorem 3:} As observed from the plots in Figure 1 related to preferential attachment graphs, the profits to the SV are greater when it accounts for externalities than when it does not, and an SV could make up to approximately 25\% extra profits (based on our model) with complete information for monopoly markets, and up to approximately 18\% in oligopoly markets (see Figure 3). This is intuitive in the sense that the SV has more user information when knowing about the externalities and can price optimally to increase its profits. However, in reality it is difficult to measure/observe the externalities. Thus, in spite of getting topological information from the insurer, an SV might have to price its products without taking externalities into account. The profits for the non-price discrimination scenario are encapsulated as a special case of $P_{1}$ when $G$ has all entries equal except the zero diagonal entries. We also observe from Figure 1 that (i) the plot trends are invariant of the graph topology, i.e., with different $\beta$ values, the sub-plots look nearly the same, and (ii) increasing the network size does not affect the plot trends either. 

Our plot results for \emph{random tree graphs} in the monopoly setting are very similar to those of preferential attachment graphs (see Figure 2), for reasons noted above except that for influence values between 0.7 and 1.0, the profit ratio curve increases towards values in the range [.95 .99] and then decreases. This trend is due to the fact that at an influence value of around 0.8, the graph topology and investment externalities have little effect on the optimal prices charged by the SV, and as a result the difference between $P_{0}$ and $P_{1}$ is minimal. Intuitively, this happens because when deciding what price to offer to a user, the monopolist considers the trade-off between profit loss due to (potentially) subsidizing the user and increase in profits due to the user's externality influence over their peers. The profit loss is proportional to the security investment of the user, and it increases with the influence of the network on this user. The profit increase term, on the other hand relates to the influence of the agent on the rest of the network. At an influence value of 0.8, the profit loss nearly equals profit gain and network and externality effects does not affect the profit ratio by a significant margin. For all other influence values, the effects of the network and investment externalities are non-signifcant. We emphasize that the rationale proposed here also applies to the trends observed on preferential attachment graphs, but there the peak point is achieved for an influence value of 1. 

\noindent \textbf{Oligopoly - Converging and Non-Converging Scenarios:} In Figures 3 and 4, we specifically see the difference in market output for two autonomous SVs. In Figure 3, we plot the profit ratios for SV1 and SV2 when we run \emph{three} iterations of the two stage pricing game involving the SVs and their clients. We see a gap in the profit ratio between the two SVs due to the non-convergence of the SV and client strategies, within three iterations. However, in Figure 4, when we plot the profit ratios under game convergence (achieved speedily, in approximately eight iterations), we observe the similarity in the profit ratios for both the SVs. The gap in the profit ratio performance between the market convergence and market  non-convergence settings is intuitive and a reflection of reality where businesses take mutiple rounds to stabilize. 

\noindent \textbf{Topology Dependent Pricing and Investments:} In the \emph{monopoly} setting, if the monopolist is only allowed to charge a uniform per-unit price to all its clients, the optimal price is computed to be $p_{0} = 1$ per unit of investment, for one instance of a 50-node (user) PA generated network with $\beta = 3$. For the same network instance we plot the heterogenous prices charged to users, in Figure 5a. \emph{We observe that the per-unit prices decrease with the increase in the centrality of nodes.}
For the two-price case, we primarily focus on obtaining the sets of users paying the discounted and regular price respectively, based on the users' security investment amounts. For a given instance of a 50-node network we first assume that the prices are given exogenously and are $p_{L} = 0.85$, and $p_{H} = 1.15$, i.e., a 15 percent deviation from the optimal single uniform price. We emphasize here that our value of 15\% is arbitrarily chosen based on our experience of binary prices of certain products in our day-to-day lives in Los Angeles. Next, we compute the optimal user investment for that same 50-node network instance, when (i) the monopolist can only use the binary pricing scheme, and (ii) when it can perfectly price discriminate. The corresponding total consumption levels for all users for the three different pricing scenarios are given in Figure 5b. 
We repeat the same process for an arbitrary instance of a 50-node graph formed using the random tree generation process noted above. Here, the optimal uniform price per client is computed to be $p_{0} = 1.2$ per unit of investment, and we again adopt the 15\% deviation rule for binary prices. We report the results in Figure 6a and 6b. For the \emph{oligopoly} setting in PA networks, we plot total consumption level for all clients of both the SVs in Figures 7a and 7b on iteratively achieving a unique subgame perfect Nash equilibria. 

The plot results suggest that for each of the three pricing scenarios in both monopoly and oligopoly markets, the resulting consumption (amount in security investments) profiles are similar. We observe that the users who are the most influential, i.e., influence the rest of the users more than they are influenced, consume the largest amounts of the good. This observation supports the theoretical result in \cite{palhui1}\cite{palhui}. Moreover, as predicted by our analysis, it is precisely these consumers that are offered the most favorable per-unit consumption prices (in the heterogenous pricing case) by the monopolist (as seen from Figures 5a and 6a.). \emph{The combination of these two observations also leads us to the fact that the total cost incurred by a network user is nearly a constant in the heterogenous pricing setting, thus leading to consumer fairness.} Finally, even when the monopolist is constrained to charging two prices, it tries to favor those central consumers, who end up getting the discounted price (see Figures 5a and 6a.). 

\noindent \textbf{Relaxing Model Assumptions:} We made a few important assumptions to simplify the model analysis that we relax while conducting a simulation study to verify the strength and necessity of the assumptions. For simulation purposes we have not assumed $Q - G$ to necessarily be positive definite, and still we observe trends of profit ratio, $\frac{P_{0}}{P_{1}}$, to follow results in Theorem 3. This implies that the assumption that $Q - G$ should be positive definite for Equation (9) to hold, is not an important one, even though from an analysis perspective, positive definiteness of the matrix $Q - G$ ensures closed form expressions of $\frac{P_{0}}{P_{1}}$ bounds. We also do not enforce $2\beta_{i} > \sum_{j\,\epsilon\,N}h_{ij}$ for simulation purposes and observe that optimal user investments are bounded for many random samples of PA and tree topologies. 

\noindent \textbf{Takeaway Message in Relation to Network Effects:} We have shown through both theory and simulations that heterogenously pricing network users based on their location and security investment amounts in a consumer overlay network is fair for each consumer (since the product of the consumption amount and the per unit price is nearly a constant for every user), and at the same time allows the SV to make more profits than it would make in case of uniform pricing (as in the current market scenario). This is true for both the monopoly as well as the oligopoly setting. The extra profit in turn allows SVs to transfer some profit to cyber-insurance agencies in a symbiotic relationship with them. 

\section{Conclusions}
In this paper, we studied a market for differentiated security product pricing, primarily with a view to ensuring that security vendors (SVs) make more profit in the differentiated pricing case compared to the case of non-differentiated pricing. We have mathematically modeled the profit made by security vendors, and proposed a novel consumer differentiated pricing mechanism for SVs based on (i) their consumers' logical network locations and (ii) security investment amounts made by the consumers. We validated our analytical model via extensive simulations conducted on practical SV client topologies, and showed that a \emph{monopoly} SV could improve their current profit margins by upto $\approx$ 25\% (based on the simulation settings) by accounting for client location in the consumer network and their investment information, whereas in an \emph{oligopoly} setting, SVs could increase their current profit margins by approximately upto $\approx$ 18\%. Specifically, the intuition behind our results (as shown via both theory and simulations) is that price discriminating consumers in proportion to the \emph{Bonacich centrality} of individual users results in maximum profit for an SV. In addition, we showed that our proposed SV pricing mechanism also ensures consumer fairness (a notion similar to network neutrality) at market equilibrium by (i) charging each consumer a per unit product usage price based on (a) their location in the logical network and (b) the amount of positive externality they generates through his security investments, and (ii) equally costing each client nearly a constant \emph{total} amount in security investments, irrespective of the client's overlay network location. Finally, we also tackled the combinatorial NP-hard problem of SVs optimally price discriminating consumers when there are only two price categories, i.e., regular and discounted. In this regard, we designed a randomized-approximation algorithm to the binary pricing problem that provides an approximation guarantee of 0.878 within the optimal solution of the total profit made by an SV.  

\section{Proofs of Theorems}
In this section we provide the proofs of Theorems 1-5. \\
 We first state and prove the relevant lemmas required for the proof of Theorem 1. \\
\textbf{Lemma 1.} \emph{The game $G^{sub}$ is supermodular\footnote{In supermodular games, the marginal utility of increasing a player's strategy increases with the increases in other players' strategies.}.} \\
\emph{Proof.} The payoff/utility functions are continuous, the strategy sets are real compact subsets, and for any two consumers $i,j\,\epsilon N$, $\frac{\partial^{2}u_{i}}{\partial x_{i}\partial x_{j}} \ge 0$. Hence $G^{sub}$ is supermodular. $\blacksquare$ \\
\textbf{Lemma 2.} The spectral radius of $Q^{-1}G$ is smaller than 1, and the matrix $I - Q^{-1}G$ is invertible. \\ 
\emph{Proof.} Let $\overrightarrow{v}$ be an eigenvector of $Q^{-1}G$ with $\lambda$ being the corresponding eigenvalue, with $|v_{i}| > |v_{j}|$ for all $j\,\epsilon\,N$. We have the following equation due to the fact that $(Q^{-1}G)\overrightarrow{v} = \lambda\overrightarrow{v}$. 
{\footnotesize
\begin{equation}
|\lambda v_{i}| = |(Q^{-1}G_{i})\overrightarrow{v}| \le \sum_{j\,\epsilon\,N}(Q^{-1}G)_{ij}|v_{j}|\le\frac{1}{2\beta_{i}}|v_{i}|\sum_{j\,\epsilon\,N}h_{ij} < \frac{v_{i}}{2}.
\end{equation}}
Here $(Q^{-1}G)_{i}$ denotes the $i-th$ row of $(Q^{-1}G)$. Since the equation holds for any eigenvalue-eigenvector pair, the spectral radius of $(Q^{-1}G)$ is strictly smaller than 1. Now observe that each eigenvalue of $I - Q^{-1}G$ can be written as $1 - \lambda$. Since the spectral radius of $Q^{-1}G$ is strictly smaller than 1, none of the eigenvalues of $I - Q^{1}G$ is zero, and thus the matrix is invertible. $\blacksquare$ \\
\emph{Proof of Theorem 1.}  Since $G^{sub}$ is a supermodular game, the equilibrium set has a minimum and a maximum element \cite{topkis}. Let $\overrightarrow{x}$ denote the maximum of the equilibrium set and let $S$ be such that $x_{i} > 0$ only if $i\,\epsilon\,S$. If $S = \phi$ there cannot be another equilibrium point, since $\overrightarrow{x} = 0$ is the maximum of the equilibrium set. Assume for a contradictory purpose that $S\ne\phi$ and there is another equilibrium $\overrightarrow{\tilde{x}}$, of the game. By the supermodularity property of $G^{sub}$, $x_{i} \ge \tilde{x}_{i}, \forall i\,\epsilon\,N$. Allow $k$ to equal $argmax_{i\,\epsilon\,N} x_{i} - \tilde{x_{i}}$. Since $\overrightarrow{x}$ and $\overrightarrow{\tilde{x}}$ are not equal, we have $x_{k} - \tilde{x_{k}} > 0$. Since at NE no consumer has an incentive to increase or decrease his consumption, we have 
\begin{equation}
x_{k} - \tilde{x}_{k}\le\frac{1}{2\beta_{k}}G_{k}(\overrightarrow{x} - \overrightarrow{\tilde{x}}) = \frac{1}{2\beta_{k}}\sum_{j}h_{kj}(x_{j} - \tilde{x_{j}}),
\end{equation}
where $G_{k}$ is the $k-th$ row of $G$. 
But we have
\begin{equation}
\frac{1}{2\beta_{k}}\sum_{j}h_{kj}(x_{j} - \tilde{x_{j}}) \le \frac{x_{k} - \tilde{x}_{k}}{2\beta_{k}}\sum_{j}h_{kj} < x_{k} - \tilde{x_{k}}
\end{equation}
Thus, we reach a contradiction and $G^{sub}$ has a unique Nash equilibrium. $\blacksquare$ \\ \\
\emph{Proof of Theorem 2.} We have from Lemma 1 that $Q-G$ is non-singular and as a result the following equation holds.
\begin{equation}
\overrightarrow{p} = \overrightarrow{\alpha} - (Q - G)\left(Q - G - \frac{G^{T} - G}{2}\right)^{-1}\frac{\overrightarrow{\alpha} - c\cdot\overrightarrow{1}}{2}
\end{equation}
Equation (4) can we rewritten as 
\begin{equation}
\overrightarrow{p} = \overrightarrow{\alpha} - \left(I - \frac{G^{T} - G}{2}(Q - G)^{-1}\right)^{-1}\frac{\overrightarrow{\alpha} - c\cdot\overrightarrow{1}}{2}
\end{equation}
By the \emph{matrix inversion lemma} \cite{cdm}, we have 
{\footnotesize
\begin{equation}
\left(I - \frac{G^{T} - G}{2}(Q - G)^{-1}\right)^{-1} = I + \frac{G^{T} - G}{2}\left(Q - \frac{G^{T} + G}{2}\right)^{-1}
\end{equation}}
Thus, from Equation (5) it follows that
\begin{equation}
\overrightarrow{p} = \frac{\alpha + c\overrightarrow{1}}{2} - \frac{G^{T} - G}{2}\left(Q - \frac{G^{T} + G}{2}\right)^{-1}\frac{\overrightarrow{\alpha} - c\overrightarrow{1}}{2}
\end{equation}
Applying Equation(7) and using the definition of weighted Bonacich centrality, we get
{\footnotesize
\[\overrightarrow{p} = \frac{\overrightarrow{\alpha} + c\cdot\overrightarrow{1}}{2} + GQ^{-1}B(G', Q^{-1}, \overrightarrow{w'}) - G^{T}Q^{-1}B(G', Q^{-1}, \overrightarrow{w'})\]}
and thus prove Theorem 2. $\blacksquare$ \\ \\
\emph{Proof of Theorem 3.} The optimal price vector of the SV without and with the consideration of externality effects are given by the following equations.
\begin{equation}
\overrightarrow{p_{0}} = \frac{\overrightarrow{\alpha} + c\cdot\overrightarrow{1}}{2}.
\end{equation}
and 
\begin{equation}
\overrightarrow{p_{1}}  = \overrightarrow{\alpha} - (Q - G)\left(Q - \frac{G + G^{T}}{2}\right)^{-1}\frac{\alpha - c\cdot\overrightarrow{1}}{2}.
\end{equation}
The corresponding consumption vectors are given by
\begin{equation}
\overrightarrow{x_{0}} = (Q - G)^{-1}\frac{\overrightarrow{\alpha} - c\cdot\overrightarrow{1}}{2}.
\end{equation}
and 
\begin{equation}
\overrightarrow{x_{1}} = (Q - G')^{-1}\frac{\overrightarrow{\alpha} - c\cdot\overrightarrow{1}}{2}.
\end{equation}
It then follows that 
\begin{equation}
P_{0} = (\overrightarrow{p_{0}} - c\cdot\overrightarrow{1})^{T}\overrightarrow{x_{0}}=\frac{\overrightarrow{\alpha} - c\cdot\overrightarrow{1}}{2}(Q - G)^{-1}\frac{\overrightarrow{\alpha} - c\cdot\overrightarrow{1}}{2}.
\end{equation}
and 
\begin{equation}
P_{1} = (\overrightarrow{p_{1}} - c\cdot\overrightarrow{1})^{T}\overrightarrow{x_{1}},
\end{equation}
$P_{1}$ can be re-written as 
\[P_{1} = X - Y,\]
where
\[X = 2\left(\frac{\overrightarrow{\alpha} - c\cdot\overrightarrow{1}}{2}\right)^{T}\left(\frac{R + R^{T}}{2}\right)^{-1}\left(\frac{\overrightarrow{\alpha} - c\cdot\overrightarrow{1}}{2}\right),\]
and
\[Y = \left(\frac{\overrightarrow{\alpha} - c\cdot\overrightarrow{1}}{2}\right)^{T}\left(\frac{R + R^{T}}{2}\right)^{-1}\left(\frac{\overrightarrow{\alpha} - c\cdot\overrightarrow{1}}{2}\right).\]
Thus, we have
\[P_{1} = \left\{\left(\frac{\overrightarrow{\alpha} - c\cdot\overrightarrow{1}}{2}\right)^{T}(Q - G')^{-1} \left(\frac{\overrightarrow{\alpha} - c\cdot\overrightarrow{1}}{2}\right)\right\},\]
Now let $\overrightarrow{v} = \frac{\alpha - c\cdot\overrightarrow{1}}{2}$. We have 
\begin{equation}
\frac{P_{1}}{P_{0}} = \frac{\overrightarrow{v}^{T}(Q - G')^{-1}\overrightarrow{v}}{\overrightarrow{v}^{T}(Q - G)^{-1}\overrightarrow{v}} \le max_{||\overrightarrow{x} = 1||}\frac{\overrightarrow{x}^{T}\left(\frac{R + R'}{2}\right)^{-1}\overrightarrow{x}}{\overrightarrow{x}^{T}\frac{R^{-1} + R^{-T}}{2}\overrightarrow{x}}.
\end{equation}
Since $\frac{R^{-T} + R^{-1}}{2}$ and $\frac{R^{T} + R}{2}$ are symmetric positive definite matrices, we have from the Rayleigh-Ritz Theorem \cite{hjr} the following.
{\scriptsize
\begin{equation}
K = \lambda_{max}\left(\left(\frac{R^{-1} + R^{-T}}{2}\right)^{-0.5}\left(\frac{R + R^{T}}{2}\right)^{-1}\left(\frac{R^{-1} + R^{-T}}{2}\right)^{-0.5}\right),
\end{equation}}
where $K = max_{||\overrightarrow{x} = 1||}\frac{\overrightarrow{x}^{T}\left(\frac{R + R'}{2}\right)^{-1}\overrightarrow{x}}{\overrightarrow{x}^{T}\frac{R^{-1} + R^{-T}}{2}\overrightarrow{x}}$.
Note that for any real matrix $A$ and invertible matrix $B$, the eigenvalues of $A$ and $B^{-1}AB$ are identical. Thus, it follows that 
\begin{equation}
K = \lambda_{max}\left(\frac{2I + RR^{T} + R^{T}R^{-1}}{4}\right)^{-1}.
\end{equation}
Now since the eigenvalues of $RR^{-T} + R^{T}R^{-1}$ are real and belong to $[-2,2]$, the eigenvalues of $\left(\frac{2I + RR^{T} + R^{T}R^{-1}}{4}\right)$ are positive, and we have
{\footnotesize
\begin{equation}
\lambda_{max}\left(\frac{2I + RR^{T} + R^{T}R^{-1}}{4}\right)^{-1} = \frac{1}{\lambda_{min}}\left(\frac{2I + RR^{T} + R^{T}R^{-1}}{4}\right).
\end{equation}}
Similarly we obtain 
\begin{equation}
max_{||\overrightarrow{x} = 1||}\frac{\overrightarrow{x}^{T}\frac{R^{-1} + R^{-T}}{2}\overrightarrow{x}}{\overrightarrow{x}^{T}\left(\frac{R + R'}{2}\right)^{-1}\overrightarrow{x}} = \lambda_{max}\left(\frac{2I + RR^{T} + R^{T}R^{-1}}{4}\right).
\end{equation}
From the above two equations it follows that 
\begin{equation}
\frac{1}{2} + \lambda_{min}\left(\frac{2I + RR^{T} + R^{T}R^{-1}}{4}\right)^{-1} \le \frac{P_{0}}{P_{1}},
\end{equation}
and
\begin{equation}
\frac{P_{0}}{P_{1}}\le \frac{1}{2} + \lambda_{max}\left(\frac{2I + RR^{T} + R^{T}R^{-1}}{4}\right)^{-1}.
\end{equation}
We have thus proved our theorem. $\blacksquare$. \\ \\
In order to prove Theorem 4, we first state the well known MAX-CUT problem \cite{gjr} is as follows:
\[max \sum_{(i,j)\,\epsilon\,E}W_{ij}(1 - x_{i}x_{j})\]
\[s.t. \,x_{i}\,\epsilon\,\{-1,+1\},\,\forall i\,\epsilon\,V,\] 
where $W$ denotes a matrix of binary weights consisting of 0s and 1s. The solution to this problem corresponds to a cut as follows: let $S$ be the agents who were assigned value 1 in the optimal solution. Then it is straightforward to see that the value of the objective function corresponds to the size of the cut defined by $S$ and $V-S$. The problem can also be re-wriiten as
\[P0: \,\,min\, \overrightarrow{x}^{T}W\overrightarrow{x}\]
\[s.t. \,x_{i}\,\epsilon\,\{-1,+1\},\,\forall i\,\epsilon\,V.\] 
Now consider the following related problem: 
\[P1: \,\,min\, \overrightarrow{x}^{T}W\overrightarrow{x}\]
\[s.t. \,x_{i}\,\epsilon\,\{-1,+1\},\,\forall i\,\epsilon\,V,\] 
where $W$ is a symmetric matrix with rational entries that satisfy $0 < W^{T} = W < 1$. In the proof of Theorem 4, we will first show that $P1$ is NP-Hard by reducing from MAX-CUT. We will then reduce P1 to OPT to claim the correctness of our theorem. \\
\textbf{Lemma 3.} \emph{P1 is NP-Hard.}\\
\emph{Proof.} We prove our claim by reducing P1 from P0. Let $W$ be the weight matrix in an instance of P0. Let $W_{\epsilon} = \frac{1}{2}(\epsilon + W)$, where $\epsilon$ is a rational number between 0 and $\frac{1}{2n^{2}}$ and $|V| = n$. Observe that for any feasible $\overrightarrow{x}$ in P0 or P1, it follows that
\[2\overrightarrow{x}^{T}W_{\epsilon}(\overrightarrow{x}) - n{2}\epsilon\le\overrightarrow{x}^{T}W\overrightarrow{x}\le 2\overrightarrow{x}^{T}W_{\epsilon}(\overrightarrow{x}) + n{2}\epsilon.\]
Because the objective of P0 is always an integer and $n^{2}\epsilon < \frac{1}{2}$, the cost of P0 for any feasible vector $\overrightarrow{x}$ can be obtained from the cost of P1 by scaling and rounding. Therefore, since P0 is NP-Hard, it follows that P1 is also NP-Hard. $\blacksquare$. \\
\emph{Proof of Theorem 4.} Having proved P1 to be NP-Hard, we now prove Theorem 4 by reducing OPT from P1. We consider special instances of OPT where $G = G^{T}$, $c = 0$, and $\overrightarrow{\alpha} = [\alpha,.....,\alpha]$, and $\alpha = p_{reg} + p_{dsc}$. OPT can then be re-casted as 
\[OPT2: \,\,min\, \overrightarrow{x}^{T}(Q-G)\overrightarrow{x}\]
\[s.t. \,x_{i}\,\epsilon\,\{-1,+1\},\,\forall i\,\epsilon\,N.\] 
Now consider an instance of P1 with $W > 0$. Note that because $x_{i}^{2} = 1$, P1 is equivalent to
\[min\, \overrightarrow{x}^{T}(W + \gamma I)\overrightarrow{x}\]
\[s.t. \,x_{i}\,\epsilon\,\{-1,+1\},\,\forall i\,\epsilon\,V,\] 
where we choose $\gamma$ as an integer such that 
\[\gamma > 4\cdot max\{\rho(W), \sum_{i,j}\frac{W_{i,j}}{min_{i,j}W_{ij}}\}\]
and $\rho(\cdot)$ is the spectral radius of its argument. The definition of $\gamma$ implies that the spectral radius of $\frac{W}{\gamma}$ is less than 1. Therefore we have 
\begin{equation}
(W + \gamma I)^{-1} = \frac{1}{\gamma}\left(I - \frac{1}{\gamma}(W - \frac{W^{2}}{\gamma}) - \frac{W^{2}}{\gamma^{3}}(W - \frac{W^{2}}{\gamma}).....\right).
\end{equation}
Since all entries of $W$ and $\left(W - (\frac{W^{2}}{\gamma})\right)$ are positive, the above equality implies that the off-diagonal entries of $(W + \gamma I)^{-1}$ are negative. Therefore $(W + \gamma I)^{-1} = (Q - G)$ for some diagonal matrix $Q$ and some $G \ge 0$. Thus, it follows that 
\begin{equation}
((Q - G)\overrightarrow{1})_{k} = \left(\frac{1}{\gamma}\left(I - \frac{W}{\gamma}+ \frac{W^{2}}{\gamma^{2}}....\right)\overrightarrow{1}\right)_{k}.
\end{equation}
Since $W > 0$, we have 
\[((Q - G)\overrightarrow{1})_{k} \ge \frac{1}{\gamma}\left(1 + \left(- \frac{W\overrightarrow{1}}{\gamma}) - \frac{W^{2}\overrightarrow{1}}{\gamma^{2}}....\right)_{k}\overrightarrow{1}\right).\]
From the definition of $\gamma$ it follows that $\frac{W\overrightarrow{1}}{\gamma} \le (\frac{(\sum_{i.j}W_{ij})}{\gamma})\overrightarrow{1}\le\frac{1}{4}\overrightarrow{1}$. The above inequality implies that 
\begin{equation}
((Q - G)\overrightarrow{1})_{k} \ge \frac{1}{\gamma}\left(1 - \frac{1}{4}\left(\sum_{l=0}^{\infty}\left(\frac{1}{4}\right)^{l}\right)\right) = \frac{1}{\gamma}\left(\frac{2}{3}\right) > 0.
\end{equation}
Thus, P1 can be reduced to an instance of OPT2 by defining $Q$ and $G$ according to $(W + \gamma I)^{-1} = (Q - G)$. Therefore it follows that OPT2, and hence OPT, are NP-Hard. $\blacksquare$.\\ \\
In order to prove Theorem 5, we first describe a semidefinite programming (SDP) relaxation for the following optimization problem: 
\[max \frac{1}{4}\sum_{i,j}w_{ij}(1 - x_{i}x_{j}))\]
subject to
\[x_{i}\,\epsilon\,\{-1, +1\}\,\forall i\,\epsilon\,V\]
This optimization problem can be reduced to 
\[max \frac{1}{4}\sum_{i,j}w_{ij}(1 - \nu_{i}\nu_{j})\]
subject to 
\[\nu_{i}\,\epsilon\,S_{n}\,\forall i\,\epsilon\,V\]
where $\nu_{i}\cdot\nu_{j}$ denotes the regular inner product of vectors $\nu_{i},\nu_{j}\,\epsilon\,\mathbb{R}^{n}$, and $S_{n}$ denotes the $n$-dimensional unit sphere, i.e., $S_{n} = \{\overrightarrow{x}\,\epsilon\,\mathbb{R}^{n}|\overrightarrow{x}\cdot \overrightarrow{x} = 1\}. $. We next show that this optimization problem leads to a semidefinite program. 

Consider the collection of vectors $\{\nu_{1},.....,\nu_{n}\}$ such that $\nu_{i}\,\epsilon\,S_{n}$. Define  a symmetric matrix $Y\,\epsilon\,\mathbb{R}^{n\times n}$ such that $Y_{ij} = \nu_{i}\nu_{j}$ and $Y_{ii} = 1$. It can be seen that $Y = F^{T}F$. where $F\,\epsilon\,R^{n\times n}$ is such that $F = [\nu_{1},.......,\nu_{n}]$. This implies that $Y\ge0$. Conversely, consider a positive semidefinite matrix $Y\,\epsilon\,\mathbb{R}^{n}$ such that $Y_{ii} = 1$. Since $Y$ is positive semidefinite, there exists $F\,\epsilon\,\mathbb{R}^{n\times n}$ (which can be obtained from Cholesky factorization of the original matrix) such that $Y = F^{T}F$, it follows that $\nu_{i}\cdot\nu_{i} = 1$. These arguments imply that the feasible set in 
\[max \frac{1}{4}\sum_{i,j}w_{ij}(1 - \nu_{i}\nu_{j})\]
subject to 
\[\nu_{i}\,\epsilon\,S_{n}\,\forall i\,\epsilon\,V\]
can be equivalently written as 
\[max \frac{1}{4}\sum_{i,j}w_{ij}(1 - \nu_{i}\nu_{j})\]
subject to 
\[Y_{i,i} = 1\,\forall i\,\epsilon\,V,\, Y\ge0\]
We now show a way to obtain a provable approximation guarantee for binary quadratic optimization problems of the form:
\[max \overrightarrow{x}^{T}Q\overrightarrow{x} + 2\overrightarrow{d}^{T}\overrightarrow{x} + z\]
subject to
\[x_{i}\,\epsilon\,\{-1,1\},\, i\,\epsilon\,\{1,....,n\},\]
where $Q$, $\overrightarrow{d}$, and $\overrightarrow{z}$ have rational entries. We observe that $\overrightarrow{x}^{T}Q\overrightarrow{x} = Trace(Q) + \overrightarrow{x}^{T}\tilde{Q}\overrightarrow{x}$, where $\tilde{Q} = Q - diag(Q)$ and $x_{i}\,\epsilon\,\{-1, +1\}$. Thus, the diagonal entries of the $Q$ matrix as part of the constant term, and we can assume that $diag(Q) = 0$ without any loss of generality. We can also assume that $Q$ is symmetric. Now consider the following optimization problem:
\[max [\overrightarrow{x};y]^{T}Q'[\overrightarrow{x};y] + z\]
subject to 
\[x_{i}\,\epsilon\,\{-1,+1\},\,i\,\epsilon\,\{1,.....,n\},\]
\[y_{i}\,\epsilon\,\{-1, +1\},\]
where $Q'$ is given by the following matrix: 
\[ Q' = \left( \begin{array}{cc}
Q & \overrightarrow{d} \\
\overrightarrow{d}^{T} & 0\end{array} \right).\]
Since $[\overrightarrow{x};y]^{T}Q'[\overrightarrow{x};y] = \overrightarrow{x}^{T}Q\overrightarrow{x} + 2y\overrightarrow{d}^{T}\overrightarrow{x}$, it follows that the optimal $\overrightarrow{x}$ and the optimal objective values of the previous two optimization problems are equal. Relaxing the previous optimization problem we get
\[max \sum_{ij}\nu_{i}\cdot \nu_{j}Q'_{ij} + z\]
subject to 
\[\nu_{i}\,\epsilon\,S_{n+1},\,i\,\epsilon\,\{1,.....,n, n+1\},\]
and obtain an equivalent semidefinite program (SDP) by defining $Y_{ij} = \nu_{i}\nu_{j}$ as follows:
\[max \sum_{ij}Y_{ij}Q'_{ij} + z\}\]
subject to 
\[Y_{ii} =1,\,i\,\epsilon\,\{1,.....,n, n+1\},\]
\[Y \ge 0.\]
Using this semidefinite relaxation one can obtain an approximate solution to the original problem. We adopt an approach used by the authors in \cite{gwir} to achieve this goal. We have the following lemma to characterize the approximate solution to our semi-definite program.  \\
\textbf{Lemma 4.} \emph{Let $z \ge \sum_{i,j}|Q'_{ij}|.$ The solution to the following optimization problem: 
\[max \overrightarrow{x}^{T}Q\overrightarrow{x} + 2\overrightarrow{d}^{T}\overrightarrow{x} + z\]
subject to
\[x_{i}\,\epsilon\,\{-1,1\},\, i\,\epsilon\,\{1,....,n\},\]
using the randomized algorithm in \cite{gwir} achieves at least 0.878 times the optimal objective value of the problem.} \\ \\
\emph{Proof.} Let $W$ denote the objective value of a solution the algorithm provides. $W_{M}$ denote the optimal solution of the underlying quadratic optimization mentioned in the lemma, and $W_{P}$ denote the optimal value of the SDP relaxation, then the corresponding optimal value can be given as 
\[W_{P} = \sum_{i,j}Q'_{ij}\nu_{i}\cdot\nu_{j} + z.\]
The solutions of the problem provided by the randomized algorithm in \cite{gwir} gives us the expected contribution of given agents $i$ and $j$ to the objective function as $Q'_{ij}\left(1 - 2\frac{arccos(\nu_{i},\nu_{j})}{\pi}\right)$. Hence, the expected value of a solution is given as 
\[E[W] = \sum_{ij}\left(1 - 2\frac{arccos(\nu_{i},\nu_{j})}{\pi}\right)Q'_{i,j} + z.\]
Now since $z \ge \sum_{i,j}|Q'_{ij}|$, it follows that both $W_{M}$ and $E[W]$ are non-negative, also since $W_{P}$ corresponds to the optimal solution of the relaxation, it follows that $W_{P} \ge W_{M}$. Using these it follows that
\[W_{P} = \sum_{i,j:Q'_{ij}>0}Q'_{ij}(1 + \nu_{i}\cdot\nu_{j}) + \sum_{i,j:Q'_{ij}<0}|Q'_{ij}(1 - \nu_{i}\cdot\nu_{j}) + z_{2}\]
and 
\[E[W] = W_{1} + W_{2}.\]
where
\[W_{1} = \sum_{i:jQ'_{ij}>0}Q'_{ij}\left(2 - 2\frac{arccos(\nu_{i},\nu_{j})}{\pi}\right)\]
and
\[W_{2} = \sum_{i,j:Q'_{ij}<0} |Q_{ij} |2\frac{arccos(\nu_{i},\nu_{j})}{\pi} + z_{2}.\]
Here $z_{2} = z - \sum_{i,j}|Q'_{ij}| > 0$. Since $\frac{arccos x}{\pi} \ge \frac{\alpha}{2}(1 - x)$ and $1 - \frac{arccos x}{\pi} \ge \frac{\alpha}{2}(1 + x)$ for all $x\,\epsilon\,[-1,+1]$, it follows that $E[W] > 0.878W_{P}\ge 0.878W_{M}$, where $\alpha = 0.878$. $\blacksquare$\\ 
A corollary of this result is 
\[E[W] + \sum_{ij}|Q'_{ij}| - z > 0.878(W_{M} + \sum_{ij}|Q'_{ij}| - z).\]
\emph{Proof of Theorem 5.} Applying the corollary from Lemma 4, the solution to the monopolist's pricing problem leads to satisfying the claims in Theorem 5. $\blacksquare$. 

\bibliography{abbrv1}

\begin{thebibliography}{10}

\bibitem{amr}
R.~Anderson and T.~Moore, ``Information security economics and beyond,'' in
  {\em Information Security Summit}, 2008.

\bibitem{leb}
M.~Lelarge and J.~Bolot, ``Economic incentives to increase security in the
  internet: The case for insurance,'' in {\em IEEE INFOCOM}, 2009.

\bibitem{lmr}
G.~A. Akerlof, ``The market for lemons - quality uncertainty and the market
  mechanism,'' {\em Quarterly Journal of Economics}, vol.~84, no.~3, 1970.

\bibitem{mthr}
M.~Thompson, ``Why cyber-insurance is the next big thing,'' in {\em CNBC
  Report}, 2014.

\bibitem{rpinfr}
R.~Pal, L.~Golubchik, K.~Psounis, and P.~Hui, ``Will cyber-insurance improve
  network security: A market analysis,'' in {\em IEEE INFOCOM}, 2014.

\bibitem{ssfw}
N.Shetty, G.Schwarz, M.Feleghyazi, and J.Walrand, ``Competitive cyber-insurance
  and internet security,'' in {\em WEIS}, 2009.

\bibitem{ballester2006s}
C.~Ballester, A.~Calv{\'o}-Armengol, and Y.~Zenou, ``Who's who in networks.
  wanted: the key player,'' {\em Econometrica}, vol.~74, no.~5, pp.~1403--1417,
  2006.

\bibitem{bramoulle2007public}
Y.~Bramoull{\'e} and R.~Kranton, ``Public goods in networks,'' {\em Journal of
  Economic Theory}, vol.~135, no.~1, pp.~478--494, 2007.

\bibitem{corbo2007importance}
J.~Corbo, A.~Calv{\'o}-Armengol, and D.~C. Parkes, ``The importance of network
  topology in local contribution games,'' in {\em Internet and Network
  Economics}, pp.~388--395, Springer, 2007.

\bibitem{galeotti2009influencing}
A.~Galeotti and S.~Goyal, ``Influencing the influencers: a theory of strategic
  diffusion,'' {\em The RAND Journal of Economics}, vol.~40, no.~3,
  pp.~509--532, 2009.

\bibitem{galeotti2010network}
A.~Galeotti, S.~Goyal, M.~O. Jackson, F.~Vega-Redondo, and L.~Yariv, ``Network
  games,'' {\em The review of economic studies}, vol.~77, no.~1, pp.~218--244,
  2010.

\bibitem{sundararajan2007local}
A.~Sundararajan, ``Local network effects and complex network structure,'' {\em
  The BE Journal of Theoretical Economics}, vol.~7, no.~1, 2007.

\bibitem{rpinfr1}
R.~Pal, L.~Golubchik, K.~Psounis, and P.~Hui, ``On a way to improve
  cyber-insurer profits: When a security vendor becomes the cyber-insurer,'' in
  {\em IFIP Networking}, 2013.

\bibitem{pc1}
Symantec, ``Personal communication,''

\bibitem{rabohme}
R.~Bohme and G.~Schwartz, ``Modeling cyber-insurance: Towards a unifying
  framework,'' in {\em WEIS}, 2010.

\bibitem{mwg}
A.~Mas-Collel, M.~D. Winston, and J.~R. Green, {\em Microeconomic Theory}.
\newblock Oxford University Press, 1995.

\bibitem{topkis}
D.~M. Topkis, {\em Supermodularity and Complementarity}.
\newblock Princeton University Press, 1998.

\bibitem{bonacich}
P.~B. Bonacich, ``Power and centrality: A family of measures,'' {\em American
  Journal of Sociology}, vol.~92, 1987.

\bibitem{palhui1}
R.~Pal and P.~Hui, ``Cyber-insurance for cyber-security: A topological take on
  modulating insurance premiums,'' {\em Performance Evaluation Review},
  vol.~40, no.~3, 2012.

\bibitem{palhui}
R.~Pal and P.~Hui, ``On differentiating cyber-insurance contracts: A
  topological perspective,'' in {\em IEEE/IFIP Internet Management Conference},
  2013.

\bibitem{gjr}
M.~R. Garey and D.~S. Johnson, {\em Computers and Intractability: A Guide to
  the Theory of NP-Completeness}.
\newblock W. H. Freeman and Company, 1979.

\bibitem{ft}
D.Fudenberg and J.Tirole, {\em Game Theory}.
\newblock MIT Press, 1991.

\bibitem{abr1}
A.~L. Barabasi and R.~Albert, ``Emergence of scaling in random networks,'' {\em
  Science}, vol.~286, 1999.

\bibitem{cdm}
C.~D. Meyer, {\em Matrix Analysis and Applied Linear Algebra}.
\newblock SIAM Press, 2000.

\bibitem{hjr}
R.~A. Horn and D.~D. Johnson, {\em Matrix Analysis}.
\newblock Cambridge University Press, 2005.

\bibitem{gwir}
M.~Goemans and D.~Williamson, ``Improved approximation algorithms for maximum
  cut satisfiability problems using sem-definite programming,'' {\em Journal of
  the ACM}, vol.~42, 1995.

\end{thebibliography}
\bibliographystyle{ieeetr}

\begin{IEEEbiography}[{\includegraphics[width=1in,height=1.0in,clip,keepaspectratio]{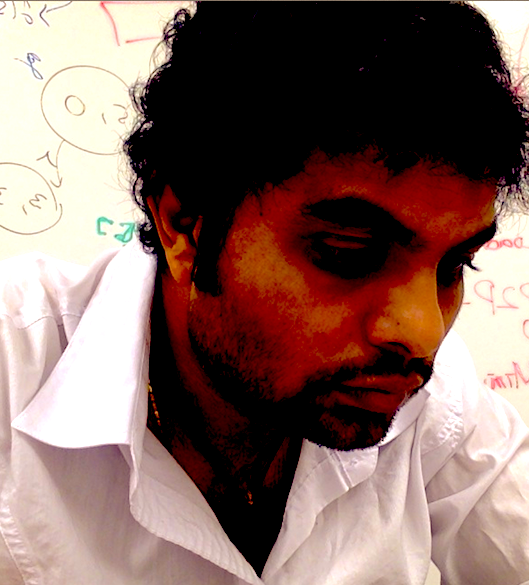}}]{Ranjan Pal}  
is a Research Scientist at the University of Southern California (USC), affiliated with both the Electrical Engineering and Computer Science departments, where he co-leads the Quantitative Evaluation and Design Group (QED). He received his PhD in Computer Science from USC in 2014, and was the recipient of the Provost Fellowship throughout his PhD studies. During his PhD, Ranjan held visiting scholar positions at the School of Engineering and Applied Science, Princeton University, USA, and Deutsch Telekom Research Laboratories (T-Labs), Berlin, Germany. His primary research interests lie in the performance modeling, analysis, and design of cyber-security, privacy, communication networks, and the Smart Grid, using tools from economics, game theory, applied probability, algorithms, information theory, and mathematical optimization. His research on cyber-insurance has generated press interests from USC News, and MIT Technology Review. Ranjan has also consulted on cyber-insurance for Accel Partners. Ranjan is a member of the IEEE and the ACM. 
\end{IEEEbiography} 

\begin{IEEEbiography}[{\includegraphics[width=1in,height=1.0in,clip,keepaspectratio]{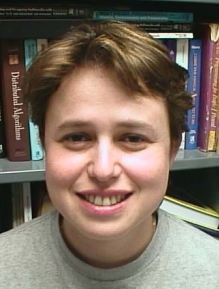}}]{Leana Golubchik} 
is a Professor in the Computer Science and EE-Systems Departments at USC.
Prior to that she was an Assistant Professor and then an Associate Professor at the University of Maryland
at College Park and an Assistant Professor at Columbia University. Leana received her PhD from UCLA in
1995. She is a past Chair (2003 - 2005) and Vice Chair (2001 - 2003) of ACM SIGMETRICS (2003 - 2005)
and a member of its Board of Directors (1999 - 2001, 2005 - 2007). She was a guest co-editor for special
issues of IEEE TKDE, the Parallel Computing Journal, and the International Journal of Intelligent Systems,
a program co-chair of the 2001 Joint ACM SIGMETRICS/Performance Conference and MIS’99, and a
program committee member of numerous conferences. Leana received several awards, including the NSF
CAREER award, the Okawa Foundation award, and the IBM and NSF Doctoral Fellowships. She is a
member of the IFIP WG 7.3 and Tau Beta Pi.
\end{IEEEbiography} 
\begin{IEEEbiography}[{\includegraphics[width=1in,height=1.0in,clip,keepaspectratio]{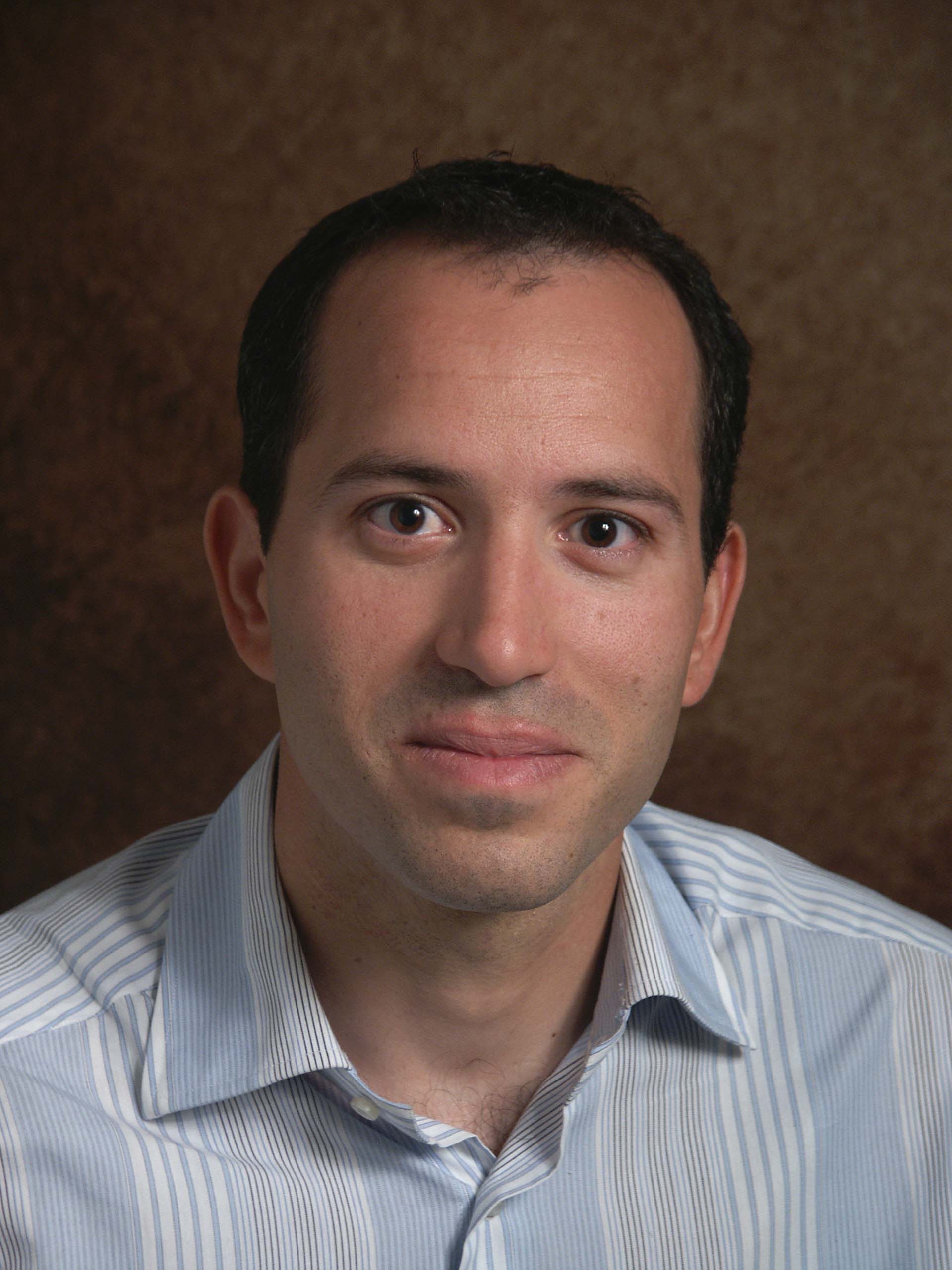}}]{Konstantinos Psounis} (M `98 - SM `08) is an
Associate Professor of Electrical Engineering and
Computer Science at the University of Southern
California. He received his first degree from the
department of Electrical and Computer Engineering
of National Technical University of Athens,
Greece, in June 1997, and the M.S.and PhD degrees
in Electrical Engineering from Stanford University,
California, in January 1999 and December 2002
respectively. Konstantinos models and analyzes the
performance of a variety of wired and wireless networks.
He also designs methods, algorithms, and protocols to solve problems
related to such systems. He is a senior member of both IEEE and ACM.
\end{IEEEbiography}
\begin{IEEEbiography}[{\includegraphics[width=1in,height=1.0in,clip,keepaspectratio]{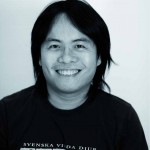}}]{Pan Hui}
is an Assistant Professor of Computer Science and Engineering at the Hong Kong University of Science and Technology. He is also a Distinguished Scientist at Deutsche Telekom Laboratories (T-Labs), Berlin. He received his PhD from Computer Laboratory,University of Cambridge. During his PhD, he was also affiliated with Intel Research Cambridge. Before that he was with University of Hong Kong for his Mphil and bachelor degree. His research interests include delay tolerant networking, mobile networking and systems, planet-scale mobility measurement, social networks, and the application of complex network science in communication system design.

\end{IEEEbiography}

\end{document}